\documentclass[]{mn2e}
\usepackage{graphicx}
\usepackage{psfig}
\usepackage{times}
\usepackage{amsmath}
\usepackage{amssymb}

\title[Oscillations of vertically integrated relativistic tori II.]
{Oscillations of vertically integrated relativistic tori --
II. Axisymmetric modes in a Kerr spacetime}

\author[Montero, Rezzolla and Yoshida]
	{Pedro J. Montero$^{(1)}$, Luciano Rezzolla$^{(1,2)}$, 
	Shin'ichirou Yoshida$^{(3,1)}$
				 			\\
							\\
	$^{(1)}$SISSA, International School for Advanced Studies and INFN,
        Via Beirut, 2 34014 Trieste, Italy		\\
	$^{(2)}$Department of Physics and Astronomy, Louisiana State
	University, Baton Rouge, LA 70803 USA 		\\
	$^{(3)}$Department of Physics, University of
	Wisconsin-Milwaukee, Milwaukee, WI 53211, USA}

\begin{document}

\maketitle
\pagerange{\pageref{firstpage}--\pageref{lastpage}}
\pubyear{2004}

\label{firstpage}

\begin{abstract}
	This  is the  second  of  a series  of  papers investigating  the
	oscillation properties  of relativistic, non-selfgravitating tori
	orbiting  around  black holes.   Extending  the  work  done in  a
	Schwarzschild  background,  we  here  consider  the  axisymmetric
	oscillations of  vertically integrated tori in  a Kerr spacetime.
	The  tori are modeled  with a  number of  different non-Keplerian
	distributions of specific angular momentum and we discuss how the
	oscillation properties depend on these and on the rotation of the
	central  black  hole.  We  first  consider  a  local analysis  to
	highlight   the   relations   between  acoustic   and   epicyclic
	oscillations  in  a Kerr  spacetime  and  subsequently perform  a
	global  eigenmode  analysis   to  compute  the  axisymmetric  $p$
	modes. In analogy with  what found in a Schwarzschild background,
	these modes behave  as sound waves that are  modified by rotation
	and   are  globally   trapped   in  the   torus.   For   constant
	distributions of specific  angular momentum, the eigenfrequencies
	appear in a sequence  2:3:4:...  which is essentially independent
	of the  size of  the disc  and of the  black hole  rotation.  For
	non-constant  distributions  of angular  momentum,  on the  other
	hand, the sequence  depends on the properties of  the disc and on
	the spin  of the black  hole, becoming harmonic  for sufficiently
	large tori.  We  also comment on how $p$  modes could explain the
	high frequency  quasi-periodic oscillations observed  in low-mass
	X-ray binaries with a black  hole candidate and the properties of
	an equivalent model in Newtonian physics.
\end{abstract}


\date{Accepted 0000 00 00.
      Received 0000 00 00.}


\section{Introduction}
\label{intro}

	A recent series of papers (Font \& Daigne, 2002a,b; Daigne \&
Font, 2004) has shown, through general relativistic hydrodynamic
simulations, that the relativistic geometrically thick discs of
high-density matter expected to form after the coalescence of a binary
system of neutron stars or after the gravitational collapse of a rotating
supermassive star may be subject to the runaway instability which could
destroy the disc on a dynamical timescale (Abramowicz, Calvani and
Nobili, 1983). Accretion tori of much smaller rest-mass densities have
also been observed in three-dimensional general relativistic
magnetohydrodynamic simulations investigating the properties of accretion
flows onto Kerr black holes (De Villiers et al. 2003).

	With the aim of assessing whether the onset of the runaway instability
depends sensitively on the choice of initial conditions, Zanotti et
al. (2003) carried out general relativistic hydrodynamic simulations
of {\it perturbed} tori orbiting around a Schwarzschild black hole and
shown that the instability may take place independently of the way the
mass accretion is induced. In addition, the simulations have indicated
how the introduction of perturbations triggers harmonic oscillations in
the tori which could produce large variations of their mass
quadrupole. In the case in which the torus is made of high-density matter
(as in the case of binary neutron star merger), these oscillations could
then lead to the emission of gravitational waves with amplitudes
comparable with those produced in a gravitational stellar-core collapse,
thus making these objects promising sources of gravitational radiation
even if the instability does not set in.

	Together with the general relativistic hydrodynamic simulations,
oscillation modes of geometrically thick relativistic discs can also be
studied through perturbative analyses. Clearly, this second approach is
computationally less intensive and allows therefore for a more detailed
investigation of the parameter space. Using a linear perturbative
approach, Rezzolla et al. (2003b; Paper I hereafter), have recently
investigated the axisymmetric oscillation properties of relativistic
tori in a Schwarzschild spacetime. To simplify the treatment and make it
as analytical as possible, the tori were built with vertically integrated
and vertically averaged quantities, thus transforming the eigenvalue
problem into a set of coupled ordinary differential equations. The tori
were then modeled with a number of different non-Keplerian distributions
of specific angular momentum.  Overall, the perturbative analysis of
Paper I confirmed the results obtained by Zanotti et al. (2003) through
nonlinear simulations, revealing that the oscillations found in the
simulations correspond to $p$ modes and that the lowest-order
eigenfrequencies must be in a sequence of small integers 2:3:4..., rather
independently of the size and the specific angular momentum distribution
of the tori. Furthermore, a detailed study of the eigenfunctions in Paper
I has also provided a possible explanation for the numerical evidence
that the runaway instability can be efficiently suppressed for tori with
distributions of specific angular momentum that follow a power-law (Font
\& Daigne, 2002b).

	This paper intends to extend the work of Paper I and is therefore
devoted to both a local and a global perturbative analysis of
axisymmetric modes of oscillation of relativistic tori in the background
spacetime of a Kerr black hole. While the local analysis provides us with
the dispersion relation for inertial-acoustic waves in relativistic
non-Keplerian discs, the global approach provides us with the
eigenfunctions and eigenfrequencies of the $p$-mode oscillations of the
system. As in Paper I, we have here removed one spatial dimension from the
problem by considering vertically integrated tori and have neglected the
variations in the background spacetime produced by the perturbations
({\it i.e.} Cowling approximation). Also in this case, therefore, the
solution of the eigenvalue problem is simplified considerably and reduces
to the solution of a single second-order ordinary differential equation.

	Overall, the results found indicate that the $p$-mode
oscillations of vertically integrated tori in a Kerr spacetime share many
of the features already encountered in a Schwarzschild spacetime. In
particular: the dependence of the fundamental eigenfrequencies on the
position of the rest-mass density maximum and on the radial size of the
discs, the relation of $p$ modes with the radial epicyclic oscillations
of point-like particles, or the simple 2:3 harmonic sequence for the
lowest-order eigenfrequencies. Furthermore, this work extends to
geometrically thick discs all of the relativistic disco-seismology
analyses carried so far for thin discs (Okazaki et al., 1987; Perez et
al., 1997; Silbergleit et al., 2001; Kato, 2001, Rodriguez et al., 2002).

	The oscillations of geometrically thick discs are important not
only because they could produce intense gravitational radiation, but also
because, in low-density discs, they may serve to explain the
high-frequency quasi-periodic oscillations (HFQPOs) observed in low-mass
X-ray binaries (LMXBs) containing a black hole candidate. In these
systems, in fact, the X-ray luminosity is modulated quasi-periodically,
giving rise to distinctive peaks in the power spectral density which, so
fare, have been found in sequences of small integers 2:3 (see Abramowicz
and Klu\'zniak, 2001 for the intepretation of the results of Remillard et
al., 1999 and Strohmayer, 2001; see Remillard et al., 2002, Homan et al.,
2003 for further observational evidences). More recently and although
with a much poorer statistics, the possible presence of a QPO structure
has been suggested also in the power spectrum of the light curve of the
two brightest X-ray flares from the Galactic Center black hole
(Aschenbach et al., 2004).

	Using the striking analogy between the results of the numerical
simulations and the observations in LMXBs, Rezzolla et al. (2003a) have
proposed a simple model that exploits the properties of $p$-mode
oscillations in thick discs and accounts for the complex phenomenology
observed for HFQPOs. It is interesting to note while such a configuration
could be produced whenever an intervening process modifies the Keplerian
character of the flow near the black hole, the numerical simulations of
De Villiers et al. (2003) have now provided a more realistic clue to how
these tori can be generated. Similarly, the recent model proposed by
Giannios \& Spruit (2004) suggests a simple way in which these $p$-mode
oscillations could be excited. Both of these models were not available at
the time the model by Rezzolla et al. (2003a) was proposed.

	To further develop this idea, we have here considered whether a
purely Newtonian description of physics could be sufficient to account
for the observations. For studying this, we have also performed a global
analysis of the axisymmetric oscillations of vertically integrated
non-Keplerian discs in Newtonian physics. Our results indicate that no
major {\it qualitative} differences emerge and that the most important
features of $p$ modes in relativistic tori remain unchanged also in the
corresponding Newtonian models, albeit with {\it quantitative}
differences.

	The plan of the paper is as follows: in Section~\ref{GRT} we
introduce the basic assumptions and equations employed in the definition
of our general relativistic, vertically integrated tori. These equations
will then be used to study axisymmetric oscillations both locally, in
Section~\ref{GRT_la}, and globally, in Section~\ref{GRT_ga}. We will
first consider configurations with constant distributions of specific
angular momentum and subsequently distributions in the cylindrical radial
coordinate distributions of specific angular momentum that increase
outwards as power-laws. In Section~\ref{QPO} we comment on how these
axisymmetric oscillations can be used to explain the HFQPOs in LMXBs
containing a black hole and discuss the solution of the eigenvalue
problem for vertically integrated Newtonian tori. Finally,
Section~\ref{conclusions} contains our conclusions.

	Hereafter, Greek indices are taken to run from 0 to 3 and Latin
indices from 1 to 3; unless stated differently, we will use units in
which $G = c = M_{\odot} = 1$.

\section{Relativistic Tori in a Kerr spacetime: Assumptions and Equations}
\label{GRT}

	We will here assume that the torus does not contribute to the
spacetime metric, which we will take to be that external to a Kerr black
hole. Furthermore, since we are interested in the portion of the
spacetime in the vicinity of the equatorial plane ({\it i.e.} for values
of the spherical angular coordinate $|\theta - \pi/2| \ll 1$), we will
write the Kerr metric in cylindrical coordinates $(t,\varpi,\phi,\theta)$
and retain the zeroth-order terms in the ratio $(z/\varpi)$. In this
case, the line element assumes the form (Novikov and Thorne 1973)
\begin{equation}
\label{metric}
ds^2 = -\frac{\varpi^2\Delta}{A}dt^2 + \frac{A}{\varpi^2}(d\phi-\omega dt)^2 
	+ \frac{\varpi^2}{\Delta}d\varpi^2 + dz^2 \ ,
\end{equation}
where $A \equiv \varpi^4+\varpi^2a^2+2M\varpi a^2$, $\Delta \equiv
\varpi^2-2M\varpi+a^2$ and $\omega \equiv 2Ma\varpi/A$. Here, $M$ is the
gravitational mass of the black hole and $a/M$ is the Kerr parameter, so
that the black hole angular momentum can be expressed as $J = aM$.

	The basic equations to be solved to construct models in
hydrostatic equilibrium are the continuity equation $\nabla_\alpha(\rho
u^\alpha)=0$ and the conservation of energy-momentum, $\nabla_\alpha
T^{\alpha\beta}=0$, where the symbol $\nabla$ refers to a covariant
derivative with respect to the metric (\ref{metric}). Here,
$T^{\alpha\beta}\equiv(e+p)u^{\alpha}u^{\beta} + p g^{\alpha \beta}$ are
the components of the stress-energy tensor of a perfect fluid, with
$u^\alpha$ being the components of the 4-velocity, $\rho$ the rest-mass
density, $e$ the energy density and $p$ the pressure. It is also useful
to introduce an orthonormal tetrad carried by the local stationary
observer and defined by the one-forms with components
\begin{eqnarray}	
{\boldsymbol \omega}^{\hat{t}} = \varpi\sqrt{\Delta/A} {\bf d}t\ ,&\quad
 	&{\boldsymbol \omega}^{\hat{\phi}} = 
  	\sqrt{A}({\bf d}\phi-\omega {\bf d}t)/\varpi\ ,
\nonumber \\ 
{\boldsymbol \omega}^{\hat{z}} = {\bf d}z\ ,&\quad
	&{\boldsymbol \omega}^{\hat{\varpi}} = {\varpi}/{\sqrt{\Delta}}
	{\bf d}\varpi\ .
\label{1-form}
\end{eqnarray}
In this frame, the components of the four velocity of the fluid are
denoted by $u^{\hat{\mu}}$ and the 3-velocity components are defined as
\begin{equation}
v^{\hat{i}}   \equiv \frac{u^{\hat{i}}}
{u^{\hat{t}}} = \frac{\omega^{\hat{i}}_{\alpha} u^{\alpha}}
	{\omega^{\hat{t}}_{\alpha}u^{\alpha}} \ , 
\hspace{1cm} i=\varpi, z, \phi \ .
\end{equation}

	We consider the perfect fluid to follow a polytropic equation of
state (EOS) $p = k \rho^{\gamma}$, where $k$ and $\gamma\equiv d\ln
p/d\ln\rho$ are the polytropic constant and the adiabatic index,
respectively. Next, being interested in a vertically-integrated
description of the torus, we introduce a {\it vertically integrated}
pressure
\begin{equation}
\label{P}
P(\varpi) \equiv \int_{-H}^{H}p dz,
\end{equation}
and a vertically integrated rest-mass density 
\begin{equation}
\label{Sigma}
\Sigma (\varpi) \equiv \int_{-H}^{H}\rho dz,
\end{equation}
where $H=H(\varpi)$ is the local ``thickness'' of the torus. We further
assume that $P$ and $\Sigma$ obey an ``effective'' polytropic EOS
\begin{equation}
\label{effctv_eos}
P = {\cal K} \Sigma^{\Gamma}\ , 
\end{equation}
so that ${\cal K}$ and $\Gamma\equiv d\ln P/d\ln\Sigma$ play the role of
the polytropic constant and of the adiabatic index, respectively. As
mentioned in Paper I, it is important to underline that
(\ref{effctv_eos}) does not represent a vertically integrated polytropic
EOS since the polytropic exponent $\Gamma=\Gamma(\varpi)$ is now a
function of the position inside the torus.

	After the vertical integration, we enforce the conditions of
hydrostatic equilibrium and axisymmetry ({\it i.e.} assume
\hbox{$\partial_t=0=\partial_{\phi}$}) and simplify the equation of
energy-momentum conservation to a Bernoulli-type form (Kozlowski et al.,
1978)
\begin{equation}
\label{bernoulli}
\frac{\nabla_i p}{e+p} = - \nabla_i \ln (u_t) +
        \frac{\Omega \nabla_i \ell}{1- \Omega \ell} \ ,
\end{equation}
where $\ell \equiv -u_{\phi}/u_t$ is the specific angular momentum ({\it
i.e.} the angular momentum per unit energy). Using the only relevant
component of equations (\ref{bernoulli}), we construct the equilibrium
model for a non-selfgravitating torus in the Kerr spacetime through the
force-balance equation
\begin{equation}
\label{vert_eq}
\frac{1}{E+P}\frac{d P}{d\varpi} 
	= - \frac{M(1-a\Omega)^2/\varpi^2-\varpi\Omega^2}
	{\varpi^2\Delta/A - A(\omega-\Omega)^2/\varpi^2} \ ,
\end{equation}
where $E$ is the vertically integrated energy density ({\it cf.}
eq. \ref{P}).

	We next perturb the hydrodynamical equations introducing Eulerian
perturbations of the hydrodynamical variables with a harmonic time
dependence of the type
\begin{equation}
\left(\delta V^{\hat{\varpi}}, \delta V^{\hat{\phi}}, \delta Q \right) \sim
	{\rm exp}({-{\rm i}\sigma t})\ ,
\end{equation}
where $\delta Q\equiv\delta P/(E+P)$ and where we have defined the
vertically averaged velocity perturbations respectively as
\begin{equation}
\label{vert_g1}
\delta V^{\hat{\varpi}} \equiv
	\frac{1}{2H}\int_{-H}^{H}\delta v^{\hat{\varpi}}dz \ , \qquad
\delta V^{\hat{\phi}} \equiv
	\frac{1}{2H}\int_{-H}^{H}\delta v^{\hat{\phi}}dz \ . 
\end{equation}
As in our previous investigation in a Schwarzschild spacetime, we assume
that the Eulerian perturbations in the metric functions can be neglected,
{\it i.e.} $\delta g_{ab}=0$ (Cowling approximation; Cowling,
1941). While this condition does not hold true in general, it represents
a very good approximation in the case of non-selfgravitating tori.

	Introducing now the quantities
\begin{equation}
\label{var_disp}
\delta U \equiv {\rm i}\delta V^{\hat{\varpi}}\ , \qquad\qquad \delta W
	\equiv \delta V^{\hat{\phi}}\ ,
\end{equation}
to eliminate the imaginary part from the system of equations and after a
bit of straightforward algebra, we derive the following set of ordinary
differential equations
\begin{eqnarray}
\label{euler-rad}
&& \hskip -0.5 cm  
\sigma\frac{\Delta}{\sqrt{A}}\delta U +
	\alpha\frac{\Delta}{\varpi^2} \delta Q'
	+ \left[\frac{\Delta^{3/2}}{A}\left(\frac{A}{\varpi^2}\right)'\Omega 
	\right. -
\nonumber\\
&&\hskip 0.5 cm
	\left. \frac{\Delta^{3/2}}{A}\left(\frac{A\omega}{\varpi^2}\right)'
	+ 2\frac{\Delta^{3/2}}{\varpi^2}(\Omega-\omega)
	\frac{P'}{E+P}\right]\delta W = 0 \ ,
\nonumber\\ 
\end{eqnarray}
\begin{eqnarray}
\label{euler-phi}
&& \hskip -0.5 cm  
\sigma\frac{\varpi^2\sqrt{\Delta}}{A}\delta W + \left\{
	\Omega' + \Omega\left[\ln\left(\frac{A}{\varpi^2}\right)'
	+\frac{A^2\omega\omega'}{\varpi^4\Delta}
	\right]\right. +
\nonumber\\
&& \hskip -0.25 cm 
	\frac{A}{\varpi^2\Delta}\left[
	\left(\frac{\varpi^2\Delta}{A}\right)'
	-\left(\frac{A\omega^2}{\varpi^2}\right)'+
	\omega\left(\frac{A\omega}{\varpi^2}\right)'
	\right] (\omega - \Omega) -
\nonumber\\
&& \hskip -0.25 cm 
	\left.\frac{A^2\omega'}{\varpi^4\Delta}\Omega^2 - 
	\frac{\varpi^{2}}{A}
	\left(\frac{A\omega}{\varpi^{2}}\right)'\right\}
	\frac{\Delta}{\sqrt{A}}\delta U
	+\frac{A\sigma \alpha}{\Delta
	\varpi^{2}}\left(\omega-\Omega\right)\delta Q = 0  \ ,
\nonumber\\
\end{eqnarray}
\begin{eqnarray}
\label{cont_gr}
&& \hskip -0.5 cm  
\sigma \delta Q + \widetilde{\Gamma}\frac{\Delta}
      	{\sqrt{A}}\delta U'
	+ \Biggl\{\frac{\Delta}{\sqrt{A}}\Biggl[\frac{P'}{E+P}+ 
\nonumber\\ 
&& \hskip 0.0cm 
	\left. 
	\widetilde{\Gamma}\left(\frac{1}{\varpi}-
	\frac{1}{2}\ln\left(\frac{r^2\Delta}{A}-
	\frac{A}{\varpi^2}(\omega-\Omega)^2
	\right)'\right)\right] +
\nonumber\\ 
&& \hskip 0.0cm 
	\widetilde{\Gamma}
	\left(\frac{\Delta}{\sqrt{A}}\right)'\Biggr\} \delta U
 	-\left(\frac{\sigma\sqrt{\Delta}(\omega-\Omega)A{\varpi^2}}
	{{\varpi^4\Delta}
	-{A^2}(\omega-\Omega)^2}\right)
	\widetilde{\Gamma}\delta W = 0 \ ,
\nonumber\\ 
\end{eqnarray}
where $\alpha \equiv 1/(u^{t})^{2}$, $\widetilde{\Gamma} \equiv {\Gamma
P}/{E+P}$, and the index $'$ indicates a radial derivative.

	Equations (\ref{euler-rad})--(\ref{cont_gr}) are the $\varpi$-
and $\phi$-components of the perturbed relativistic Euler equations and
the perturbed continuity equation, respectively. As in Paper I, they are
solved numerically for the eigenfrequencies and eigenfunctions of
$p$-mode oscillations of an oscillating vertically integrated thick disc
in a Kerr spacetime. Before discussing the solution of this eigenvalue
problem, however, it is instructive to consider how wave-like
perturbations propagate in the torus.

\section{perturbations of relativistic tori: a local analysis}
\label{GRT_la}

	We now present a local analysis of the perturbed hydrodynamical
equations (\ref{euler-rad})--(\ref{cont_gr}) which will serve to clarify
the relation between acoustic waves and other waves that play a
fundamental role in fluids orbiting in a central potential, {\it i.e.}
inertial (or epicyclic) waves. For this analysis we assume a harmonic
radial dependence in the perturbed fluid quantities of the type $(\delta
Q, \delta U, \delta W) \sim {\rm exp(i}k\varpi)$, where $k$ is the radial
wavenumber so that the local wavelength is $\lambda=2 \pi/k$.

	To make the system of equations simpler, we have removed from
equations (\ref{euler-rad})--(\ref{cont_gr}) those terms that are much
smaller than the others. Details on these simplifications are discussed
in the Appendix~\ref{appendix}. The linearized perturbation equations
(\ref{euler-rad})--(\ref{cont_gr}) resulting from this procedure can thus
be written as a homogeneous linear system in matrix form
\begin{equation}
\label{lin_sys}
{\bf M}
\left(k,\sigma\right)
	\left(\begin{array}{c}
	  \delta U \\ \delta W \\ \delta Q\end{array}\right)
	= 0 \ ,
\end{equation}
where ${\bf M}$ is the coefficient matrix given by
\begin{equation}
{\bf M}(k,\sigma)= 
\nonumber
\end{equation}
\begin{eqnarray}
\label{emme}
&& \hskip -0.75 cm
	\left(
	\hskip -0.3cm
	\begin{array}{ccc}
	\hskip -0.1cm
	\displaystyle
	\frac{\sigma \Delta}{\sqrt{A}} &
	\hskip -0.1cm
	\displaystyle
	\frac{\Delta^{\frac{3}{2}}}{A}
	\left[
	\Omega\left(\frac{A}{r^{2}}\right)'
	-\left(\frac{A\omega}{r^{2}}\right)'+
	\frac{2A\left(\Omega-\omega \right)P'}{r^{2}(E+P)}\right]&
	\hskip -0.1cm
	\displaystyle
	\frac{{\rm i}k\alpha\Delta}{r^{2}} \\ \\
	\hskip -0.1cm
	\displaystyle
	\frac{H \Delta}{\sqrt{A}} &
	\displaystyle
	\frac{\sigma r^{2} \sqrt{\Delta}}{A} & 0 \\ \\
	\hskip -0.0cm
	\displaystyle
	\frac{{\rm i}k \widetilde{\Gamma}\Delta}{\sqrt{A}} & 0 & \sigma
	\end{array}
	\hskip -0.3cm
	\right)
\nonumber 
\hskip -1.5 cm  \\
\end{eqnarray}

	The dispersion relation is then obtained by imposing the
determinant of ${\bf M}$ to be zero, thus guaranteeing that a non-trivial
solution to the system of equations (\ref{lin_sys}) exists. After a bit
of algebra one obtains the dispersion relation
\begin{equation}
\label{disp_rel}
\sigma^{2} = \kappa_{\rm r}^{2} + 
	\frac{\Delta}{\varpi^{2}} \left[ \frac{\varpi^{2}\Delta}{A} -
	\frac{A(\omega^{2}+\Omega^{2})}{\varpi^{2}}\right]
	k^{2} c_{s}^{2}\ ,
\end{equation}
where $c_{s}\equiv \sqrt{dP/dE}$ is the relativistic sound speed within
the vertically integrated disc and where the relativistic radial
epicyclic frequency for an extended fluid object in a Kerr spacetime
$\kappa_{\rm r}$ is defined as
\begin{eqnarray}
\label{kr}
&&
\hskip -0.4cm
\kappa_{\rm r}^{2} \equiv \frac{\Delta C}{\varpi^{2}}
	\left[\left(\frac{A}{\varpi^{2}}\right)'\Omega-
	\left(\frac{A\omega}{\varpi^{2}}\right)'+
	\frac{2A}{\varpi^{2}}\left(\Omega-\omega\right)
	\frac{P'}{E+P}\right] \ .
\nonumber \\
\end{eqnarray}

\noindent Here, the quantity $C$ is just a shorthand for 
\begin{eqnarray}
\label{C}
&& 
\hskip -0.75 cm
C\equiv 
	\frac{A(\omega-\Omega)}{\varpi^{2}\Delta}
	\left[\left(\frac{\varpi^{2}\Delta}{A}\right)'-
	\left(\frac{A\omega^{2}}{\varpi^{2}}\right)'+
	\omega\left(\frac{A\omega}{\varpi^{2}}\right)'\right]
	+
\nonumber\\
&& \hskip -0.5cm	
	\Omega\left[\ln\left(\frac{A}{\varpi^{2}}\right)'+
	\frac{A^{2}\omega\omega'}{\Delta
	\varpi^{4}} -\frac{\Omega A^{2}\omega'}{\Delta \varpi^{4}} 
	\right]  
	+\Omega' \hfill
	- \frac{\varpi^{2}}{A}\left(\frac{A\omega}{\varpi^{2}}\right)'
	\ .
\nonumber \\
\end{eqnarray}

	Equations (\ref{disp_rel}) and (\ref{kr}) include corrections
coming from the rotation of the black hole and pressure gradients from
the fluid distribution and reduce respectively to the dispersion relation
and to the relativistic epicyclic frequency for an extended object
obtained in Paper I for a Schwarzschild black hole ({\it cf.} eqs. 42 and
43 of Paper I). When the motion is almost Keplerian, the pressure
gradients are negligible and expression (\ref{kr}) reduces to the
expression for the radial epicyclic frequency for point-like particles
(Okazaki et al., 1987)
\begin{equation}
\label{okazaki}
\left(\kappa_{\rm r}^{2}\right)_{\rm Kep} =
	\frac{M}{\varpi^3}
	\left[
	\frac{1 - 6M/\varpi \pm 8a \sqrt{M/\varpi^3} -3a^{2}/\varpi^2}
	{(1 \pm 8a \sqrt{M/\varpi^3})^2}
	\right]\ ,
\end{equation}
where the $\pm$ sign distinguishes orbits that are corotating from those
that counter-rotating relative to the black hole spin.

\begin{figure}
\centering
\includegraphics[angle=0,width=8.5cm]{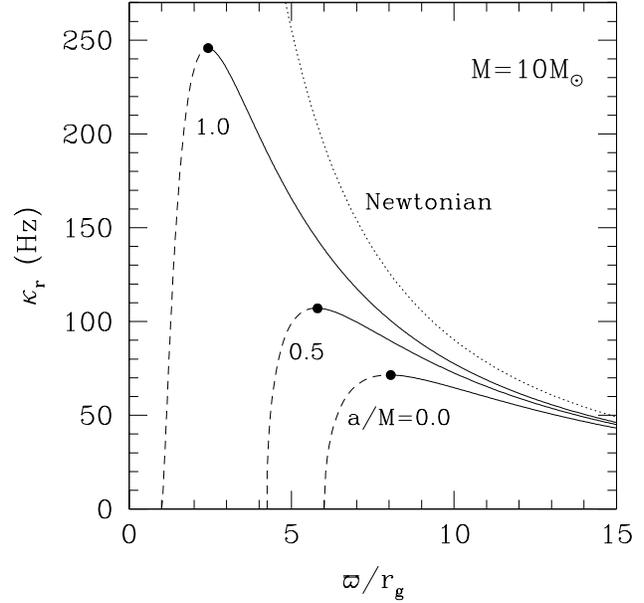}
\caption{Radial epicyclic frequency for different values of the black
hole spin. In each curve, a dashed line refers to the ``increasing''
branch ({\it i.e.} with $\kappa'_{\rm r} > 0$), while a solid line to the
``decreasing'' branch ({\it i.e.} with $\kappa'_{\rm r} < 0$). Indicated
with solid circles are the locations of the maximum frequencies for each
black hole spin, while the dotted line refers to the Newtonian values for
the radial epicyclic frequency. The numerical values have been computed
for a black hole with mass $M=10M_{\odot}$ and the radial extents have
been expressed in units of gravitational radii $r_g \equiv GM/c^2$.}
\label{epic} 
\end{figure}

\begin{figure}
\centering
\includegraphics[angle=0,width=8.5cm]{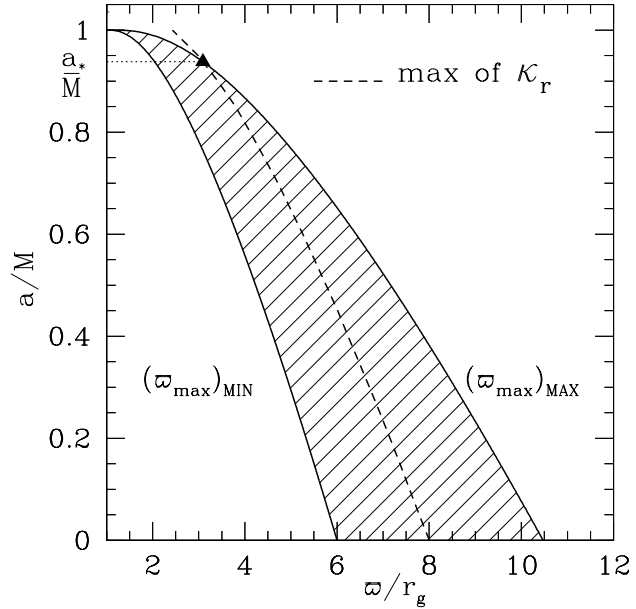}
\caption{Possible locations of the rest-mass density maximum $\varpi_{\rm
max}$ for tori with constant specific angular momentum distribution and
for all possible black hole spins. The solid lines refer to the smallest
and largest possible positions of $\varpi_{\rm max}$, while the dashed
line indicates the locations of the maximum of epicyclic frequency
$\kappa_{_r}$ ({\it cf.} the solid circles in Fig.~\ref{epic}).}
\label{rmax_vs_a} 
\end{figure}

	The behaviour of the radial epicyclic frequency (\ref{okazaki})
is shown in Fig.~\ref{epic} where different curves refer to different
black hole spins and where we consider corotating orbits only.  For each
curve, furthermore, a dashed line refers to the ``increasing'' branch of
the epicyclic frequency ({\it i.e.} with $\kappa'_{\rm r} > 0$), while a
solid line refers to the ``decreasing'' branch ({\it i.e.} with $\kappa'_{\rm r}
< 0$). The solid circles indicate the locations of the maximum
frequencies for each black hole spin and will be used subsequently in
Fig.~\ref{rmax_vs_a}. Finally, the dotted line indicates the
Newtonian radial epicyclic frequency which will be used for the
discussion in Section~\ref{ga}. The numerical values refer to a black
hole with mass $M=10M_{\odot}$ and the radial extents have been expressed
in units of gravitational radii $r_g \equiv GM/c^2$.

	Overall, the dispersion relation (\ref{disp_rel}) shows that the
propagation of small perturbations in a fluid rotating around a Kerr
black hole is characterized by two main features. The first one is of
purely acoustic nature ({\it i.e.} with $\sigma \propto k c_s$); the
second of is of purely inertial nature ({\it i.e.} with $\sigma \propto
\kappa_{\rm r}$) and is reminiscent of the oscillations that a particle
orbiting in a central potential experiences when perturbed. For these
reasons, the resulting waves are usually referred to as {\it
inertial-acoustic waves} and the corresponding modes can be classified as
$p$ modes. The determination of the eigenfrequencies and eigenfunctions
of these modes will be discussed in the following Section.

\begin{figure}
\centering
\includegraphics[angle=0,width=8.5cm]{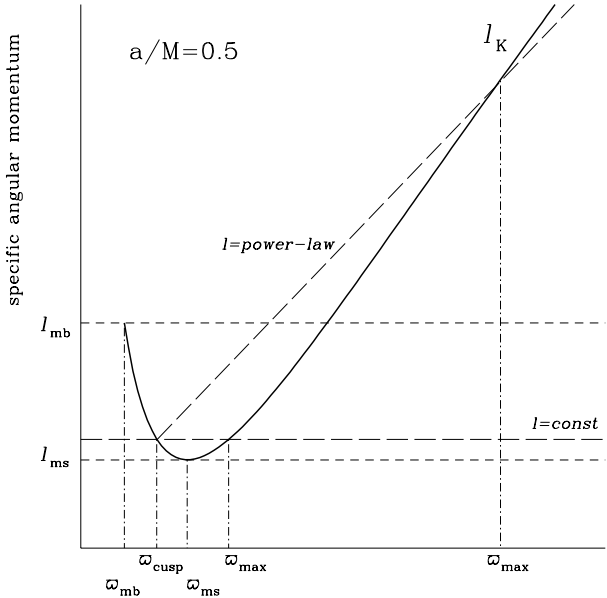}
\caption{Schematic representation of the position of the relevant radii
in the cases of  ``constant'' and ``power-law'' distributions of
specific angular momentum. In particular, while $\varpi_{\rm cusp}$ and
$\varpi_{\rm max}$ indicate the positions of the cusp and of the
rest-mass density maximum, $\varpi_{\rm mb}$ and $\varpi_{\rm ms}$
indicate the positions of the marginally bound and marginally stable
orbits.}
\label{l_scheme} 
\end{figure}

\section{perturbations of relativistic tori: a global analysis}
\label{GRT_ga}

	We now turn to a global analysis of the axisymmetric oscillation
of relativistic tori in a Kerr spacetime by solving the system of
equations (\ref{euler-rad})--(\ref{cont_gr}) as an eigenvalue problem. In
particular, the solution is found  using a ``shooting'' method
(Press et al., 1986) in which, once the appropriate boundary conditions
are provided, two trial solutions are found, starting from inner and
outer edges of the disc respectively, and these are then matched 
at an intermediate point where the Wronskian
of the two solutions is evaluated (see Paper I for details). This
procedure is iterated until a zero of the Wronskian is found, thus
providing a value for $\sigma$ and a solution for $\delta Q, \delta U$,
and $\delta W$. The numerical methods employed here to solve the
eigenvalue problem are the same as those discussed in Paper I, where a
more detailed discussion can be found.

	It is useful to briefly summarize how the background model for
the torus is constructed (see also Paper I) as this will help clarifying
the discussion of the results presented in the following Sections. First
of all, a value of the black hole spin $a/M$ is chosen and the
distribution of the specific angular momentum $\ell=\ell(\varpi)$ is
selected.  The positions of the cusp and of the maximum rest-mass density
in the torus are then obtained by imposing that the specific angular
momentum at these two points coincides with the Keplerian value (see
Fig.~\ref{l_scheme} for a schematic view). The inner edge of the torus
$\varpi_{\rm in}$ is determined by fixing the potential gap, $\Delta
W_{in}=W_{\rm in}-W_{\rm cusp}$, defined as
\begin{equation}
\label{potentilagap}
\Delta W_{\rm in}=\ln[(-u_{t})_{\rm in}] - \ln[(-u_{t})_{\rm cusp}]-
	\int_{\ell_{\rm cusp}}^{\ell_{\rm in}}
	\frac{\Omega d\ell}{1-\Omega \ell}\ ,
\end{equation}
with $\Delta W_{\rm in}=0$ corresponding to a torus filling its outermost
equipotential surface. Next, the hydrostatic balance equation
(\ref{vert_eq}) is integrated from $\varpi_{\rm in}$ to the outer edge of
the torus $\varpi_{\rm out}$, defined as the position at which
$P=0$. Sequences of tori having different radial extents for a given
distribution of specific angular momentum and black hole spin can then be
constructed by varying the potential gap $\Delta W_{in}$. In such
sequences, the tori will have the same rest-density maxima $\varpi_{\rm
max}$.

	Clearly, different distributions of specific angular momentum
will produce tori with different positions of the cusp and of the maximum
rest-mass density. In order to investigate how the axisymmetric
oscillations depend on these properties, we have constructed models with
different distributions of specific angular momentum, taken to be either
constant within the torus, or increasing outwards according to a generic
power-law in radius ({\it cf.}  eq. \ref{powerlaw}). The main properties
of the various models considered are summarized in Table~\ref{tab1},
where they are presented in terms of dimensionless quantities. A simple
conversion of the dimensionless eigenfrequency $\widetilde{\sigma}$ and
of the specific angular momentum $\widetilde{\ell}$ is obtained, for
instance, through the following relations
\begin{equation}
\sigma = \widetilde{\sigma}\left(\frac{M}{M_{\odot}}\right)^{-1}
	\left(\frac{GM_{\odot}}{c^{2}}\right)^{-1}c\ ,
\end{equation}
and
\begin{equation}
\ell = \widetilde{\ell}\left(\frac{M}{M_{\odot}}\right)
	\left(\frac{GM_{\odot}}{c^{2}}\right)c\ .
\end{equation}

\subsection{Constant specific angular momentum tori}
\label{lconst}

\subsubsection{Fixed black hole spin}

	We first consider tori with a distribution of specific angular
momentum that is constant in space, {\it i.e.} $\ell(\varpi)={\rm
const.}$, not only because this choice is mathematically simpler to
investigate, but also because it allows for a straightforward
determination of the permitted range of equilibrium models. In this case,
in fact, for a torus of finite size to exist, the value of the specific
angular momentum must satisfy \hbox{$\ell_{\rm ms} < \ell < \ell_{\rm
mb}$}, where $\ell_{\rm ms}$ and $\ell_{\rm mb}$ are the specific angular
momenta of the marginally stable and of the marginally bound orbit,
respectively. For a Schwarzschild black hole, $\ell_{\rm ms} =
3\sqrt{6}/2 \sim 3.67$ and $\ell_{\rm mb} = 4$, while they depend of the
the black hole spin for a Kerr black hole. Because the position of the
rest-mass density maximum $\varpi_{\rm max}$ will depend on the specific
value of $\ell$ chosen, it is then possible to define the range in which
$\varpi_{\rm max}$ lies. In other words, all of the possible finite-size
tori will have $(\varpi_{\rm max})_{_{\rm MIN}} < \varpi_{\rm max} <
(\varpi_{\rm max})_{_{\rm MAX}}$, where we have defined $(\varpi_{\rm
max})_{_{\rm MIN}}$ as the value of $\varpi_{\rm max}$ when
$\ell=\ell_{\rm ms}$, and $(\varpi_{\rm max})_{_{\rm MAX}}$ as the value
of $\varpi_{\rm max}$ when $\ell=\ell_{\rm mb}$. Clearly, $(\varpi_{\rm
max})_{_{\rm MIN}}$ also coincides with the location of the marginally
stable orbit, while $(\varpi_{\rm max})_{_{\rm MAX}}$ provides the
position of the rest-mass density maximum for a torus whose outermost
equipotential surface closes at infinity.

\begin{table*}
\begin{center}
\caption{Main properties of the equilibrium models studied. From left to
right the columns show: the black hole spin parameter,the type of
specific angular momentum distribution, the constant coefficient
$\ell_{c}$ and the power-law index $q$ ({\it cf.} eq. \ref{powerlaw}),
the position of the maximum density point $\widetilde{\varpi}_{\rm max}$,
the position of the inner and outer radii of the torus
$\widetilde{\varpi}_{\rm in}$ and $\widetilde{\varpi}_{\rm out}$, the
position of the maximum density point with respect to the epicyclic
frequency curve, and the potential gap at the inner edge of the torus
$\Delta W_{\rm in}$. This latter quantity is reported because not all
equilibrium models reported in this table have been constructed as the ones filling the largest
closed equipotential surface ({\it i.e.} Pl6--Pl9).}

\label{tab1}
\begin{tabular}{l|c|c|c|c|c|c|c|c|c|c}
\hline Model & $a/M$ &${\rm \ell(\varpi)-distribution}$ & $\rm
{\widetilde{\ell_{c}}}$ & $q$ & $\widetilde{\varpi}_{\rm max}$ &
$\widetilde{\varpi}_{\rm in}$ & $\widetilde{\varpi}_{\rm out}$ & $L$
&$\kappa'_{\rm r}$\ \ at\ \ ${\widetilde{\varpi}_{\rm max}}$ & $\Delta
W_{\rm in}$ \\

\hline
C1a    & 0.5 & const.& 3.283 & 0.0 & 6.01 & 3.21 & 13.26 & 10.04 & $<0$ & $-1.0 \times 10^{-10}$ \\
C1b    & 0.5 & const.& 3.313 & 0.0 & 6.29 & 3.13 & 18.04 & 14.92 & $<0$ & $-1.0 \times 10^{-10}$ \\ 
C1c    & 0.5 & const.& 3.343 & 0.0 & 6.55 & 3.05 & 26.66 & 23.60 & $<0$ & $-1.0 \times 10^{-10}$ \\   
C1d    & 0.5 & const.& 3.373 & 0.0 & 6.81 & 2.99 & 47.78 & 44.79 & $<0$ & $-1.0 \times 10^{-10}$ \\
C1e    & 0.5 & const.& 3.403 & 0.0 & 7.07 & 2.93 & 181.6 & 178.7 & $<0$ & $-1.0 \times 10^{-10}$ \\
\hline
C2a    & 0.5 & const.& 3.193 & 0.0 & 5.00 & 3.65 & 6.330 & 2.680 & $>0$ & $-1.0 \times 10^{-10}$ \\ 
C2b    & 0.5 & const.& 3.207 & 0.0 & 5.20 & 3.54 & 7.128 & 3.583 & $>0$ & $-1.0 \times 10^{-10}$ \\ 
C2c    & 0.5 & const.& 3.223 & 0.0 & 5.39 & 3.45 & 8.079 & 4.627 & $>0$ & $-1.0 \times 10^{-10}$ \\ 
C2d    & 0.5 & const.& 3.253 & 0.0 & 5.71 & 3.32 & 10.22 & 6.896 & $>0$ & $-1.0 \times 10^{-10}$ \\ 
\hline 
C3a    & 0.0 & const.& 3.704 & 0.0 & 7.0  & 5.22 & 8.572 & 3.354 & $<0$ & $-1.0 \times 10^{-10}$ \\ 
C3b    & 0.1 & const.& 3.637 & 0.0 & 7.0  & 4.71 & 9.566 & 4.851 & $<0$ & $-1.0 \times 10^{-10}$ \\ 
C3c    & 0.2 & const.& 3.573 & 0.0 & 7.0  & 4.24 & 11.18 & 6.946 & $<0$ & $-1.0 \times 10^{-10}$ \\ 
C3d    & 0.3 & const.& 3.511 & 0.0 & 7.0  & 3.79 & 14.27 & 10.48 & $<0$ & $-1.0 \times 10^{-10}$ \\ 
C3e    & 0.4 & const.& 3.452 & 0.0 & 7.0  & 3.36 & 22.40 & 19.04 & $<0$ & $-1.0 \times 10^{-10}$ \\ 
C3f    & 0.5 & const.& 3.395 & 0.0 & 7.0  & 2.95 & 103.1 & 100.2 & $<0$ & $-1.0 \times 10^{-10}$ \\ 
\hline 
C4a    & 0.5 & const.& 3.167 & 0.0 & 4.5 & 3.99 & 4.821 &  0.828 & $<0$ & $-1.0 \times 10^{-10}$ \\ 
C4b    & 0.6 & const.& 3.060 & 0.0 & 4.5 & 3.32 & 5.650 &  2.330 & $<0$ & $-1.0 \times 10^{-10}$ \\ 
C4c    & 0.7 & const.& 2.960 & 0.0 & 4.5 & 2.70 & 7.879 &  5.173 & $<0$ & $-1.0 \times 10^{-10}$ \\ 
C4d    & 0.8 & const.& 2.867 & 0.0 & 4.5 & 2.14 & 32.55 &  30.39 & $<0$ & $-1.0 \times 10^{-10}$ \\ 
\hline 
Pl0    & 0.0 & power-law & 3.381 & 0.06 & 9.0 & 5.02 & 15.92 & 10.90 & $<0$ & $-1.0 \times 10^{-10}$ \\ 
Pl1    & 0.1 & power-law & 3.120 & 0.09 & 9.0 & 4.97 & 16.04 & 11.08 & $<0$ & $-1.0 \times 10^{-10}$ \\ 
Pl2    & 0.2 & power-law & 2.999 & 0.10 & 9.0 & 4.64 & 18.61 & 13.97 & $<0$ & $-1.0 \times 10^{-10}$ \\ 
Pl3    & 0.3 & power-law & 2.800 & 0.13 & 9.0 & 4.49 & 19.81 & 15.32 & $<0$ & $-1.0 \times 10^{-10}$ \\ 
Pl4    & 0.4 & power-law & 2.738 & 0.13 & 9.0 & 4.04 & 26.52 & 22.48 & $<0$ & $-1.0 \times 10^{-10}$ \\ 
Pl5    & 0.5 & power-law & 2.730 & 0.13 & 9.0 & 3.49 & 66.39 & 62.89 & $<0$ & $-1.0 \times 10^{-10}$ \\ 
Pl6    & 0.6 & power-law & 2.649 & 0.14 & 9.0 & 3.74 & 197.7 & 194.0 & $<0$ & $-1.0 \times 10^{-2}$  \\
Pl7    & 0.7 & power-law & 2.362 & 0.19 & 9.0 & 3.71 & 289.4 & 285.7 & $<0$ & $-1.0 \times 10^{-2}$  \\
Pl8    & 0.8 & power-law & 2.272 & 0.20 & 9.0 & 4.84 & 29.12 & 24.28 & $<0$ & $-5.0 \times 10^{-2}$  \\
Pl9    & 0.9 & power-law & 2.152 & 0.22 & 9.0 & 4.45 & 46.05 & 41.60 & $<0$ & $-1.0 \times 10^{-1}$  \\
\hline 

\end{tabular}
\end{center}
\end{table*}

	Furthermore, because in a Kerr spacetime the values of $\ell_{\rm
ms}$ and $\ell_{\rm mb}$ will depend on the black hole spin $a/M$, it is
possible to define the set of all the possible values that $\varpi_{\rm
max}$ can assume for constant angular momentum tori construncted in a
Kerr spacetime. This is shown as a shaded area in Fig.~\ref{rmax_vs_a},
where we have also plotted with solid lines the different values of
$(\varpi_{\rm max})_{_{\rm MIN}}$ and $(\varpi_{\rm max})_{_{\rm MAX}}$
as the black hole spin is increased from 0 to 1
\footnote{We have not considered here counterotating black holes, {\it
i.e.} with $a<0$.}. Also shown with a dashed line are the locations of
the maximum of epicyclic frequency for different black hole spins ({\it
cf.} the solid circles in Fig.~\ref{epic}).

	Note that for tori with constant specific angular momentum
distributions around rapidly rotating black holes with spin parameter
$a/M$ larger than $a_{*}/M$ ({\it cf.} Fig. \ref{rmax_vs_a}), all of the
possible values of $\varpi_{\rm max}$ will be located on the ``increasing
branch'' of the epicyclic frequency ({\it i.e.} at positions where
$\kappa'_{\rm r} > 0$). As a result, the closer the maximum density point
is to the black hole horizon, the smaller the value of the epicyclic
frequency at that point ({\it cf.} eq. \ref{okazaki}). For rotations of
the central black hole smaller than $a_{*}/M$, on the other hand, the
models with $\varpi_{\rm max}$ in the ``decreasing branch'' of the
epicyclic frequency ({\it i.e.}  at positions where $\kappa'_{\rm r} <
0$) are also allowed together with those on the increasing branch. In
this case, if the maximum rest-mass density point coincides with the
location of the maximum point of the epicyclic frequency ({\it i.e.}  at
positions where $\kappa'_{\rm r} = 0$), the eigenfrequency of the
fundamental mode for a given radial extent will have the largest value
for any $a < a_*$. The locus of these points is indicated with a dashed
line in Fig.~\ref{rmax_vs_a}.

\begin{figure*}
\centering
\includegraphics[angle=0,width=8.5cm]{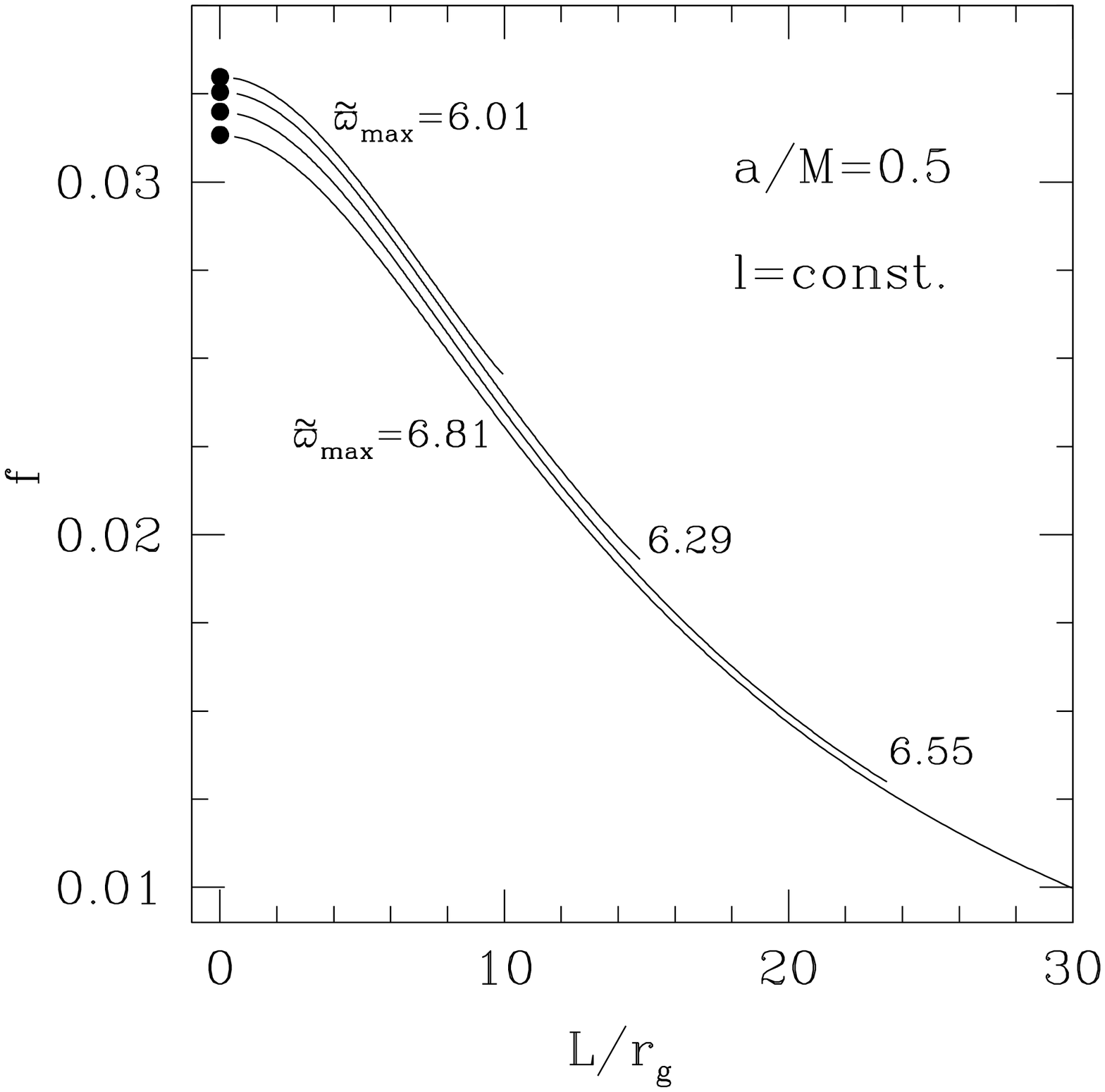}
\hskip 0.5 cm
\includegraphics[angle=0,width=8.5cm]{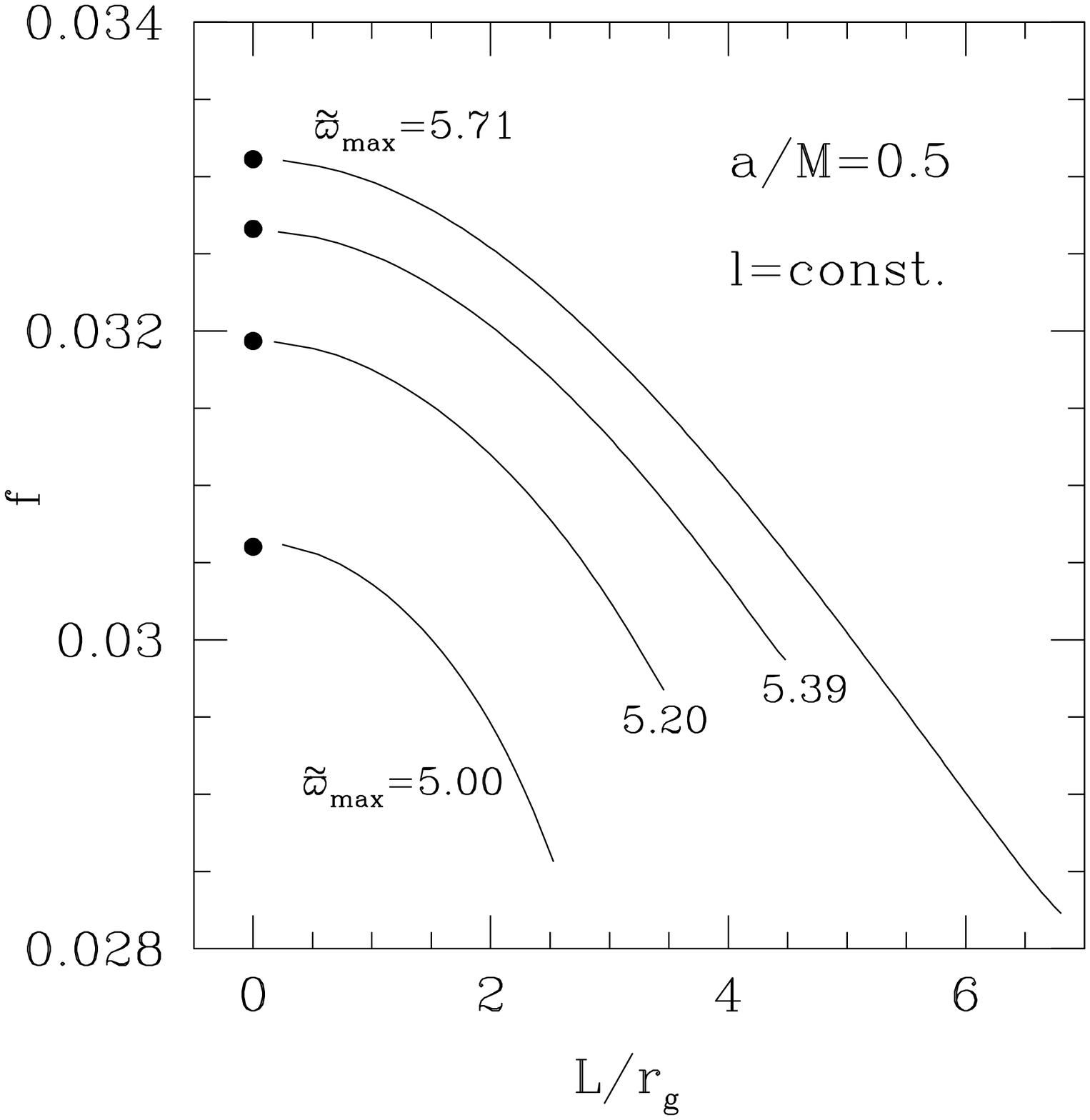}
\caption{Eigenfrequencies for the fundamental mode of axisymmetric $p$
modes for tori with a constant distribution of specific angular
momentum. Each line corresponds to a sequence of tori having the same
$\varpi_{\rm max}$ but different radial extents $L$ and the solid circles
correspond to the values of the Keplerian radial epicyclic frequency
(\ref{okazaki}) at $\varpi_{\rm max}$. Both panels refer to a Kerr black
hole with $a/M=0.5$, but while the right panel has $\varpi_{\rm max}$
located in the increasing branch of the epicyclic frequency ({\it i.e.}
$\kappa'_{\rm r}>0$), the left one has $\varpi_{\rm max}$ located in the
decreasing branch of the epicyclic frequency ({\it i.e.} $\kappa'_{\rm
r}<0$). }
\label{a_fixed} 
\end{figure*}

\begin{figure*}
\centering
\includegraphics[angle=0,width=8.5cm]{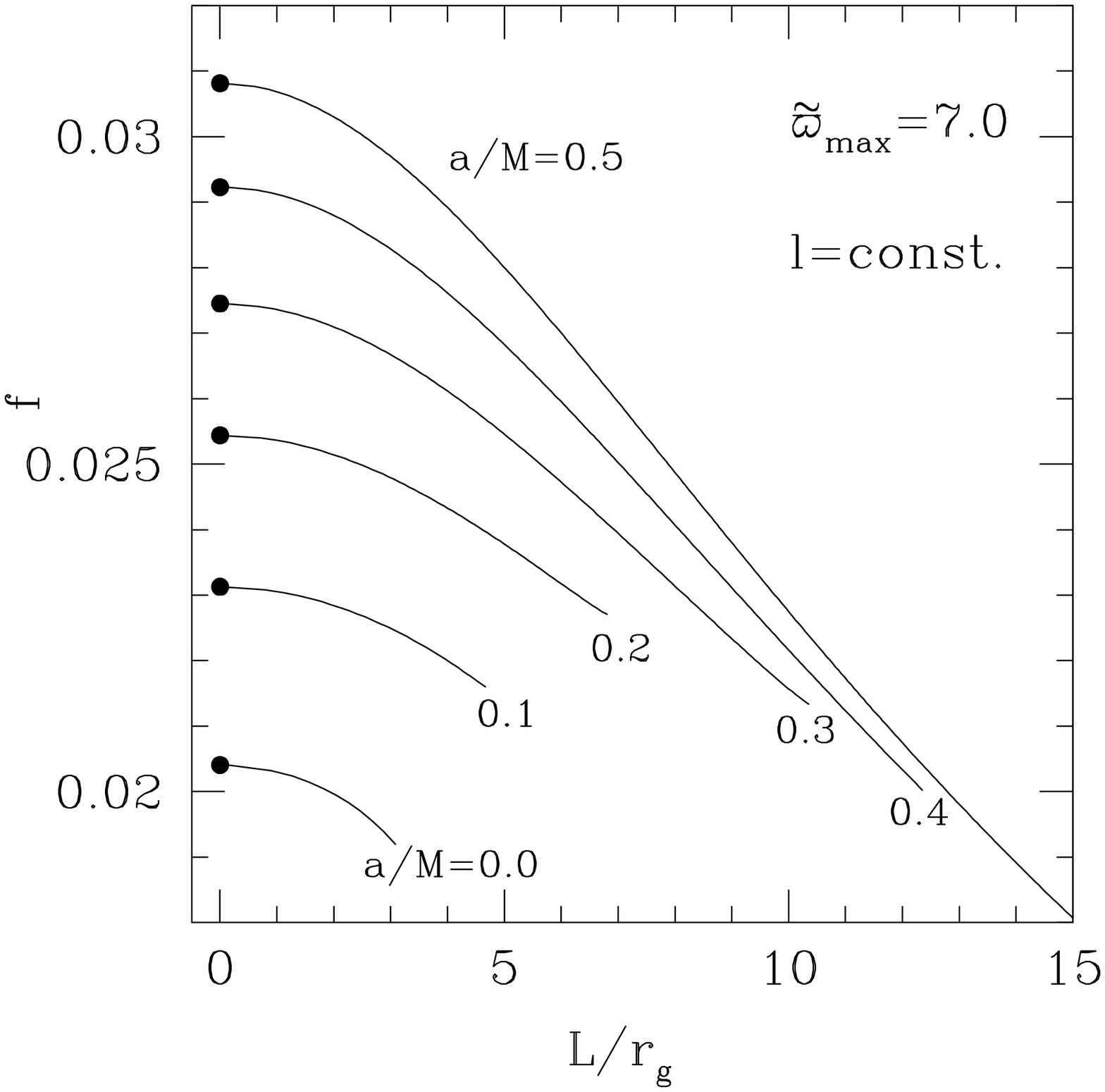}
\hskip 0.5 cm
\includegraphics[angle=0,width=8.5cm]{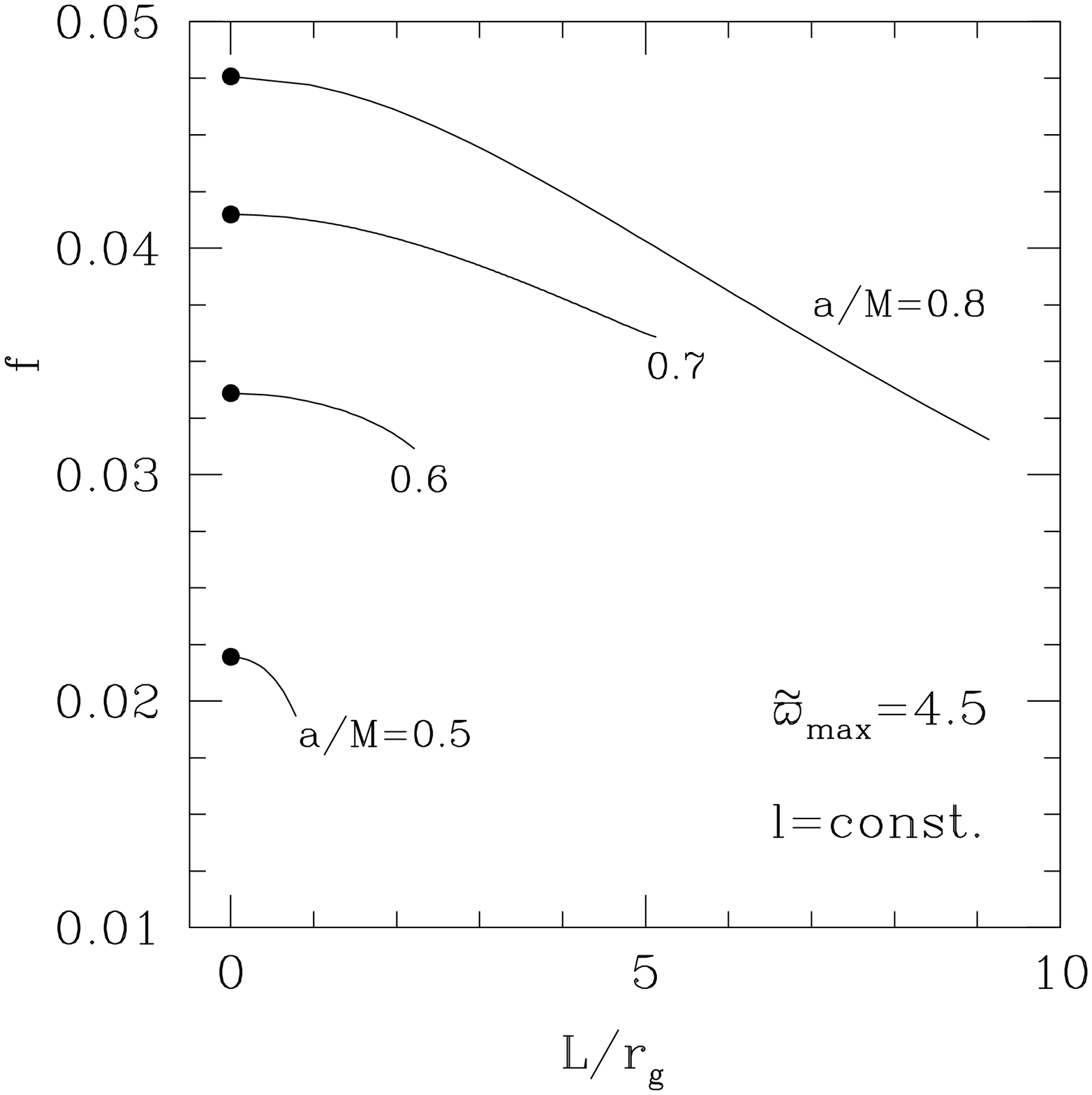}
\caption{Eigenfrequencies for the fundamental mode of axisymmetric $p$
modes for tori with a constant distribution of specific angular
momentum. Each line refers to a sequence of tori for the same black hole
spin $a/M$, but having different radial extents $L$ and the solid circles
correspond to the values of the Keplerian radial epicyclic frequency
(\ref{okazaki}) at $\varpi_{\rm max}$. In each panel, the sequences have
been computed keeping fixed the position of $\varpi_{\rm max}$, which are
however different according to whether the black holes are slowly
rotating (left panel) or rapidly rotating (right panel). }
\label{wmax_fixed} 
\end{figure*}

	In order to investigate the dependence of the fundamental mode of
oscillation on the position of the maximum rest-mass density, we have
first considered sequences of tori having different radial extents $L
\equiv \varpi_{\rm out} - \varpi_{\rm in}$. All sequences are built for
the same value of the black hole spin, while each sequence is
characterized by a specific value of $\varpi_{\rm max}$, which is chosen
to be either on the increasing or on the decreasing branch of the
corresponding epicyclic frequency. The results for a black hole spin
$a/M=0.5$ are summarized in the two panels of Fig.~\ref{a_fixed}, which
show the values of the eigenfrequency of the fundamental mode of
oscillation $f$ as a function of the disc radial extents $L$.

	In the left panel, in particular, we show the results for models
whose maximum rest-mass density is located on the decreasing branch of
the radial epicyclic frequency. As expected for modes behaving
effectively as sound waves trapped in a rotating fluid, the
eigenfrequencies decrease like $L^{-1}$ as the radial extent of the torus
increases. Note also that, as was shown in Paper I for tori orbiting
around a Schwarzschild black hole, the eigenfrequencies of the
fundamental mode tend to the values of the radial epicyclic frequency at
$\varpi_{\rm max}$ as the radial dimension of the discs tends to zero (In
Figs.~\ref{a_fixed}, \ref{wmax_fixed} and \ref{newt} the filled circles
show the values of the Keplerian radial epicyclic frequency \ref{okazaki}
at $\varpi_{\rm max}$.). These results confirm the expectation that, as
their size diminishes, the discs effectively behave as rings of particles
in circular orbits and have as their characteristic frequency the
epicyclic frequency at the maximum rest-mass density point.

	Note also in the left panel of Fig.~\ref{a_fixed} that models
with the same radial extent and different constant angular momentum
distributions, {\it i.e.}, tori with different locations of $\varpi_{\rm
max}$, have fundamental eigenfrequencies that become larger as the
rest-mass density maximum approaches the black hole. This behaviour can
be explained simply since, for $L \to 0$, the fundamental
eigenfrequencies converge to the values of the epicyclic frequencies and
the latter increase for smaller radii as one moves along the decreasing
branch.

	The opposite behavior is observed in the right panel of
Fig.~\ref{a_fixed}, but in this case the maximum density point of all
models is located on the increasing branch of the epicyclic frequency. As
a result, models with the same radial extent have fundamental frequencies
that become smaller as the rest-mass density maximum approaches the black
hole. Therefore, for any given radial extent, the model with the largest
fundamental-mode eigenfrequency will be the one with $\varpi_{\rm max}$
located at the maximum of the epicyclic frequency ({\it i.e.} along the
dashed line in Fig.~\ref{rmax_vs_a}).

\subsubsection{Influence of the black hole spin}

	In order to investigate the influence of the spin of the central
black hole on the axisymmetric oscillations, we have solved the
eigenvalue problem for sequences of tori having the same location of
$\varpi_{\rm max}$ but different black hole rotation rates.

	We note that, in the case of constant specific angular momentum,
it is not possible to construct sequences of tori having the same
location of the rest-mass density maximum for all the possible values of
the black hole spin. This is due to the fact that for any given
$\varpi_{\rm max}$, the existence of marginally stable and 
marginally bound orbit limits the range of values that $a/M$ can assume.
This is quite apparent in Fig.~\ref{rmax_vs_a} where, for any chosen
value of $\varpi_{\rm max}$, the black hole spin cannot span the whole
range from $0$ to $1$.

	Because of this constraint, we show in the left panel of
Fig.~\ref{wmax_fixed} sequences of tori having the same value of
$\varpi_{\rm max}/M=7.0$ and black hole spins varying from $a/M=0.0$ to
$a/M=0.5$. Clearly, these sequences share many of the properties
discussed for Fig.~\ref{a_fixed} and, in addition, for any given radial
size of the torus, the frequency of the fundamental mode will be larger
for larger black hole spin.  Indeed, this is what is expected, since the
radial epicyclic frequency for a point-like particle $(\kappa_{\rm
r})_{_{\rm Kep}}$ increases monotonically with $a$ for any value of
$\varpi$ ({\it cf.}  Fig.~\ref{epic}).  The same behaviour is observed
also when $\varpi_{\rm max}/M=4.5$, which allows for larger rotations
rates of the black hole and is shown in the right panel of
Fig.~\ref{wmax_fixed}.

\subsubsection{Harmonic Sequence}

	An interesting and potentially important feature of axisymmetric
$p$-mode oscillations of tori with constant specific angular momentum is
that the lowest-order eigenfrequencies appear in a sequence 2:3:4...,
with a good accuracy ({\it i.e.} within 10\%) and rather
independently of the details of the disc properties. This feature was
first discovered through the fully nonlinear numerical simulations of
Zanotti et al. (2003) and subsequently confirmed through a perturbative
analysis in a Schwarzschild spacetime in Paper I, and through additional
independent numerical simulations (Klu\'zniak et al., 2004; Lee et al.,
2004).

\begin{figure}
\centering
\includegraphics[angle=0,width=8.5cm]{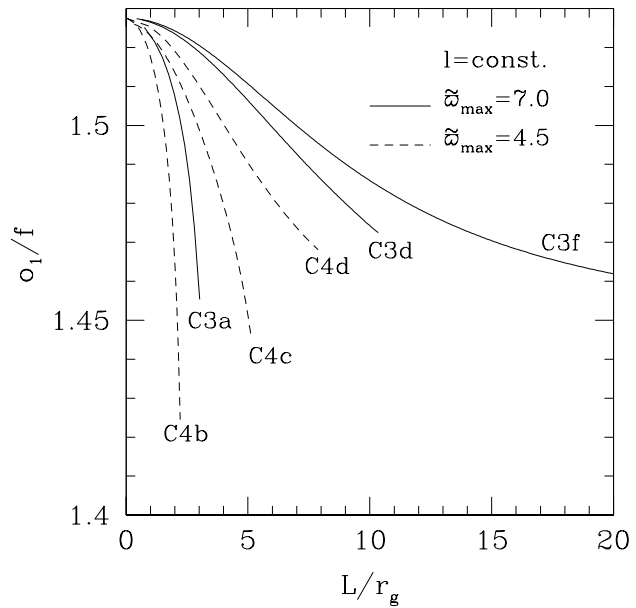}
\caption{Ratio of the first overtone and the fundamental mode of
oscillation for tori with constant specific angular momentum
distributions. The different sequences refer to the different models
described in Table~\ref{tab1}, with the solid lines referring to
$\varpi_{\rm max}=7.0\, r_g$ and the dashed ones to $\varpi_{\rm
max}=4.5\, r_g$.}
\label{o1_lconst} 
\end{figure}

       A harmonic relation for the lowest-order eigenfrequencies continues 
to exist also for tori with constant specific angular momentum in a
Kerr spacetime and rather insensitively of
the black hole spin. Fig.~\ref{o1_lconst} shows the ratio between the
frequencies of the first overtone ${o}_1$ and the fundamental mode ${f}$ for some representative
tori with constant specific angular momentum distributions considered in
Table~\ref{tab1}. Different line types correspond to sequences that have
been constructed with different locations of the rest-mass density
maximum. All of the sequences span values of $a/M$ from 0 to 0.8 without
showing major qualitative differences.

	Since it is not the result of a mathematical constraint but,
rather, the consequence of a global mode of oscillation, the 2:3 ratio
between the fundamental mode and its first overtone is satisfied with an
accuracy that is in general within 10\%, but is, nevertheless, not
exact. A number of different elements can contribute to a deviation from
an exact relation among integers and the results of our calculations show
that conditions such as the size of the disc, the location of the
rest-mass density maximum, the black hole spin, the distribution of
specific angular momentum, or the EOS considered, can all influence this
departure. Note also that in Fig.~\ref{o1_lconst} some of the models
({\it e.g.}  C4b and C3a) appear to have a rather large variation in the
ratio $o_1/f$. It should be recalled, however, that these tori exist only
over a very small range in the radial extent $L$ (either because of the
high value of the specific angular momentum or because of the large spin
of the black hole). Being trapped in such small tori rather close to
the black hole horizon where the angular velocity is larger, the $p$-mode
oscillations will be particularly sensitive on variations of the radial
size.

	Another interesting feature of the lowest-order eigenfrequencies
is that, for any given black hole rotation speed, their ratio tends to a single
value ${o}_1/{f}\sim 1.52$ as the size of the tori tends to zero ({\it
cf.} Fig.~\ref{o1_lconst}). This is in good qualitative agreement with
what was found in the perturbative analysis of slender tori (Blaes 1985),
where the dispersion relation for generic perturbations of Newtonian
slender tori having constant distributions of specific angular momentum, was
derived ({\it cf.}  eq. 1.8 of Blaes 1985, with $j=1$, and $|k|=0$).

\subsection{Non-constant specific angular momentum tori}
\label{lnconst}
 
	We next investigate axisymmetric $p$-modes in tori with
non-constant distributions of specific angular momentum. Clearly, in this
case the problem is highly degenerate since there is a complete
arbitrariness on what is the most appropriate distribution of angular
momentum. The first guide in the infinite possible choices is that they
should correspond to stable fluid configurations. While in Newtonian
theory one can make use of the simple Rayleigh stability criterion for
rotating inviscid fluids $d \ell /d\varpi > 0$ (Tassoul 1978), the
situation in a general relativistic framework is more complex. In
particular, the condition of stability requires that, for the perfect
fluids considered here, the gradient of the specific angular momentum
must never point toward the interior of the quasi-circular level surfaces
of $\ell$ (Seguin 1975, Abramowicz and Prasanna 1985). Because of this,
we have used a simple prescription for the specific angular momentum in
which it increases outwards with a power-law distribution of the type
\begin{equation}
\label{powerlaw}
\widetilde{\ell} = \ell_{c}\varpi^{q} \ ,
\end{equation}    
where both $\ell_c$ and $q$ are positive constants ({\it cf.}
Table~\ref{tab1} for the different values used).

	Following the procedure presented in the previous Section, we
have constructed tori by selecting the angular momentum distribution such
that the position of the cusp is always located between the marginally
bound and the marginally stable orbits. Note that, in principle, this is
not the only possible choice. Non-constant distributions of specific
angular momentum, in fact, allow for the construction of tori having the
cusp at radii larger than the position of the marginally stable orbit. In
such tori, matter lost through the cusp as a result of small
perturbations would be able to find new stable circular orbits and, as a
result, the accretion onto the black hole will require mechanisms other
than simple perturbations in the flow. Here, we have not investigated the
properties of these discs whose dynamics, however, has been considered in
detail by Zanotti et al., (2004).

	Overall, the solutions for tori with non-constant distributions
of specific angular momentum share qualitative similarities with those
found when $\ell={\rm const.}$ In particular, the main properties of the
$p$-mode oscillations in this case can be summarized as follows: Firstly,
the eigenfrequencies depend on the location of the rest-mass density
maximum $\varpi_{\rm max}$ and on whether the latter is located on the
increasing or decreasing branch of the radial epicyclic
frequency. Secondly, all eigenfrequencies tend to the radial epicyclic
frequency at the location of $\varpi_{\rm max}$ in the limit of small
tori. Thirdly, as the result of the presence of an {\it evanescent-wave}
region (see Paper I for a detailed discussion), the eigenfunctions become
vanishingly small at the inner edge of the disc, while they do not do so at
the outer edge. Finally, for any sequence built with fixed $\varpi_{\rm
max}$, the eigenfrequencies increase as the spin of the black hole is
increased.

\begin{figure}
\centering
\includegraphics[angle=0,width=8.5cm]{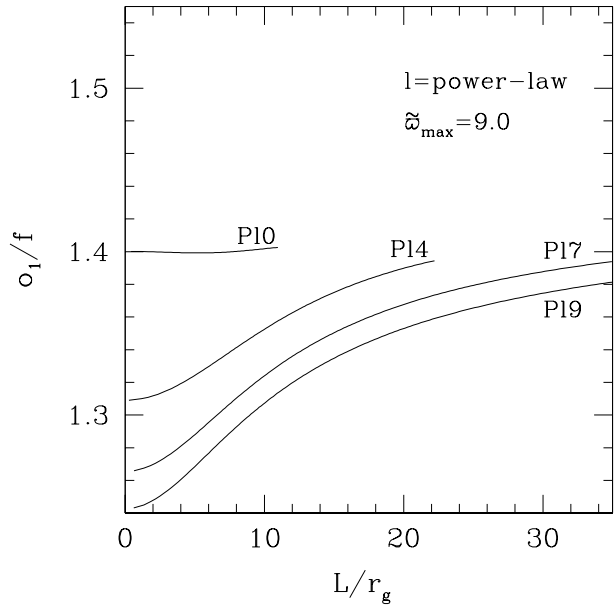}
\caption{Ratio of the first overtone and the fundamental mode of
oscillation for tori with a power-law distribution of specific angular
momentum. The different sequences distinguish different choices of the
index $q$, but all refer to the same value for $\ell_c$ ({\it cf.}
Table~\ref{tab1}).}
\label{o1_lpower2} 
\end{figure}

	Despite these similarities, a difference does emerge when
investigating non-constant angular momentum tori. In this case, in fact,
the ratio between the first overtone and the fundamental frequency of
oscillation $o_1/f$ has a behaviour which is more complex than the one
found when $\ell={\rm const.}$ In particular, it depends not only to the
position of $\varpi_{\rm max}$, but also on the values of $\ell_c$ and
$q$, as well as on $a/M$. Most importantly, while the
eigenfrequencies remain in a harmonic sequence for sufficiently large
discs, departures from a 2:3 ratio can be observed for very small
discs. This is shown in Fig.\ref{o1_lpower2}, where the ${o}_1/{f}$ ratio
is plotted for sequences of models orbiting around central black holes
rotating at different rates and having different distributions of
specific angular momentum (see Table~\ref{tab1} for the model
descriptions). In all the cases considered, however, the departure from a
2:3 ratio is less than 20\%.

\begin{figure}
\centering
\includegraphics[angle=0,width=8.5cm]{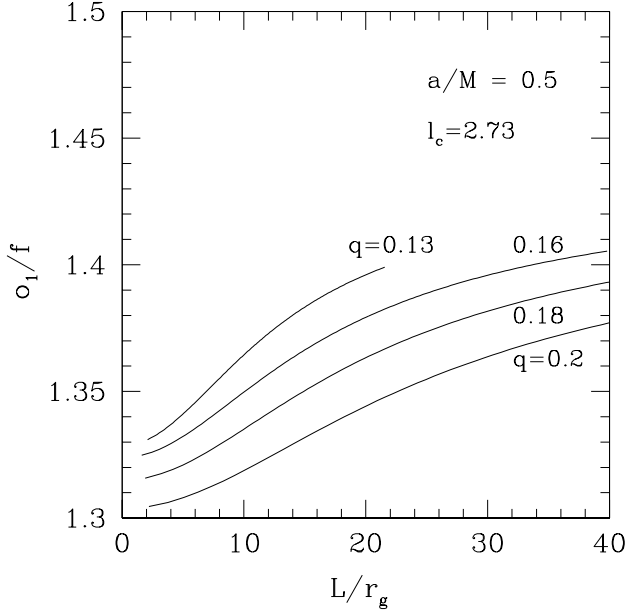}
\caption{Ratio of the first overtone to the fundamental mode shown as a
function of the radial extent of the torus $L$. The tori are modelled with a
power-law distribution of specific angular momentum and the different lines
refer to different choices of the index $q$, but all assume the same
value of $\ell_{c}=2.73$.}
\label{o1_lpower} 
\end{figure}

	Finally, note that, as the size of the discs tends to zero, the
ratios do not tend to the same value, as was the case for constant
specific angular momentum tori but, rather, to a value that depends on
the index $q$ of the power-law. This is shown in Fig.~\ref{o1_lpower},
where we illustrate results for sequences of tori with the same $\ell_{c}$
but different indices $q$, built around the same Kerr black hole with
$a/M=0.5$. Also in this case, a closer agreement with a harmonic relation
between the lowest-order eigenfrequencies is recovered for sufficiently
large discs.

\section{Implications for HFQPO{\small s} in LMXB{\small s}}
\label{QPO}

	Observations of X-ray emissions from binary systems comprising a
stellar-mass compact object have long been considered important
tools to test General Relativity in strong-field regimes. In
particular, the high frequency quasi-periodic oscillations 
(HFQPOs) observed in LMXBs binaries containing a black hole candidate
 have been proposed as a tool for measuring black hole properties such as  mass
and spin in a direct way. In such systems, the X-ray observations show a
particularly rich phenomenology in which the luminosity is modulated
quasi-periodically, giving rise to distinctive peaks in the power
spectral density. A puzzling feature of the QPO frequencies in these
systems is that they are found in sequences of small integers ({\it i.e.}
1:2, 2:3, or 1:2:3; Abramowicz and Klu\'zniak, 2001; Remillard et al.,
2002; Homan et al., 2003).

	Numerous models have been suggested to explain the HFQPOs and
their peculiar harmonic property (see Abramowicz and Klu\'zniak, 2004 for
a recent review) but the scarcity of the observational data, on the one
hand, and the crudeness of the different models, on the other hand, have
not yet made it possible to determine which model provides the most
likely description of the processes responsible for the QPOs. It should
be noted the first steps towards a more quantitative and physically
realistic description of the emission properties of possible QPO models
have recently been taken (Schittman and Bertschinger, 2004; Schnittman,
2004).

	Among the proposed models, one is particularly simple as it is
based on the assumption that the accretion disc around the black hole
terminates with a sub-Keplerian part, {\it i.e} a torus of small size
(Rezzolla et al., 2003a). This model was motivated by the  numerical
simulations of relativistic tori in a Schwarzschild spacetime (Zanotti et
al., 2003) and by the subsequent perturbative study of $p$-mode
oscillations of relativistic vertically integrated tori presented in
Paper I. A Cornerstone of this model is the evidence, both numerical and
analytic, that in these objects the fundamental mode and the first
overtones are found to be in the harmonic sequence 2:3:4... to a good
precision and in a very wide parameter space.

        Because the $p$ modes considered in the above model represent the
basic oscillation modes of a non-Keplerian disc, some of their features
are not expected to be very sensitive to the properties of the spacetime
which determine the motion of the fluid. The harmonic 2:3 relation
between the fundamental frequency and the first overtone is one  such
feature and, indeed, this property has been encountered essentially
unmodified in Schwarzschild and Kerr spacetimes, but also in a Newtonian
gravitational potential. Because of this, it is then natural to consider
whether a fully Newtonian description of the physics is sufficient to account
for the phenomenology observed in HFQPOs, or whether General Relativity
is indeed necessary. This question in addressed in the following Section.

\subsection{Can Newtonian Physics explain the HFQPOs?}
\label{ga}

	In order to assess whether $p$-mode oscillations in purely
Newtonian tori could account for the HFQPO phenomenology, we have
performed a global analysis of the axisymmetric oscillations of
vertically integrated non-Keplerian discs in Newtonian physics and
compared the results with those obtained in a general relativistic
framework.

	Consider therefore a non-selfgravitating perfect fluid torus in
equilibrium and orbiting a central object of mass $M_{0}$ producing a
gravitational potential $\Psi$. As in the general relativistic case, the
system of equations describing the torus will be simplified by
considering vertically integrated or vertically averaged quantities,
which we indicate as $P$ and $\Sigma$, in analogy with the corresponding
relativistic quantities ({\it cf.} eqs. \ref{P} and \ref{Sigma}). In
order to handle simpler expressions we normalize all quantities as
\begin{equation}
\label{norm1}	
\widetilde{v}^{\,i} \equiv \frac{v^{i}}{c_{0}}\ , \qquad 
	{\widetilde{\Sigma}}\equiv \frac{\Sigma}{\Sigma_{0}}\ , \qquad
	{\widetilde{P}}\equiv \frac{P}{P_{0}}\ , \qquad
	{\widetilde{\varpi}}\equiv \frac{\varpi}{\varpi_{0}}
	\ , 
\end{equation}
where $i=\varpi,\phi$, while $\Sigma_{0}$ is the maximum mass density and
where we have defined
\begin{equation}
\label{norm2}	
P_{0}\equiv K\Sigma_{0}^{1+1/N}\ , \qquad
	c^{2}_{0}\equiv \left(1+\frac{1}{N}\right)\frac{P_{0}}{\Sigma_{0}}\ ,  
	\qquad 
	\varpi_{0}\equiv \frac{GM_{0}}{c^{2}_{0}}\ .
\end{equation}

	The equilibrium configurations are then obtained by solving the
hydrostatic equilibrium equation which, in terms of the normalized
quantities, is written simply as
\begin{equation}
\frac{1}{{\widetilde{\Sigma}}}\widetilde{P}'= 
	{\widetilde{\Omega}}^{2}{\widetilde{\varpi}} - 
	{\widetilde{\Psi}}'\ ,
\end{equation}
where ${\widetilde{\Psi}}\equiv {\Psi}/{c^{2}_0}$ and
${\widetilde{\Omega}}\equiv \Omega/\Omega_0$, with $\Omega_0$ being the
angular velocity at $\varpi_0$. Introducing now the Emden function
${\widetilde{\Theta}}$ and its related quantity ${\widetilde{Q}} \equiv
N{\widetilde{\Theta}}$, the normalized rest-mass density
${\widetilde{\Sigma}}$ can written as ${\widetilde{\Sigma}} =
{\widetilde{\Theta}}^{N}$, and the equation of hydrostatic equilibrium is
expressed as an ordinary differential equation for $\widetilde{Q}$, {\it
i.e.}
\begin{equation}
\label{hydro_eq}
{\widetilde{Q}}'= 
	\widetilde{\Omega}^{2}\widetilde{\varpi} - 
	\frac{1}{\widetilde{\varpi}^{2}}\ .
\end{equation}
Once the angular velocity distribution $\Omega=\Omega(\varpi)$ is fixed
and a choice is made for the inner radius ${\widetilde \varpi}_{\rm in}$,
equation (\ref{hydro_eq}) can be integrated to provide the background
Newtonian torus in equilibrium.

	For simplicity we now restrict our attention to constant specific
angular momentum distributions so that the equation of hydrostatic
equilibrium (\ref{hydro_eq}) simplifies to
\begin{equation}
\label{hydro_eq_lconst}
{\widetilde{Q}}'= 
	\frac{\ell^2}{\widetilde{\varpi}^3} - 
	\frac{1}{\widetilde {\varpi}^2}= 
	-\left(\frac{\ell^2}{2\widetilde{\varpi}^2} - 
	\frac{1}{\widetilde {\varpi}} \right)'
	\ .
\end{equation}
Before discussing the solution of the eigenvalue problem, it is useful to
recall the features of the Newtonian models that represent important
differences from the relativistic counterparts. The first one is that the
right-hand side of equation (\ref{hydro_eq_lconst}) is just the
derivative of an effective potential ({\it i.e.} the one in the round
brackets) and it is therefore straightforward to calculate the inner edge
of the largest possible torus as the radial position at which this
potential vanishes, {\it i.e.}  $({\widetilde \varpi}_{\rm in})_{_{\rm
MIN}}=\ell^2/2$.

	The second feature to notice is again easily derived from
equation (\ref{hydro_eq_lconst}) and shows that the position of the mass
density maximum ({\it i.e.} $Q'=0$) is simply given by ${\widetilde
\varpi}_{\rm max} = \ell^2$, coinciding with the position at which the
specific angular momentum assumes a Keplerian value. As in relativistic
tori, ${\widetilde \varpi}_{\rm max}$ grows with increasing specific
angular momentum; unlike relativistic tori, however, no lower or upper
limits exist for ${\widetilde \varpi}_{\rm max}$. Indeed, for vanishingly
small values of $\ell$, it is possible to construct tori whose pressure
maxima lie arbitrarily close to the rotation axis. This property is the
consequence of the fact that, in Newtonian physics and for constant
distributions of specific angular momentum, $\ell$ is not constrained to
lie in a finite interval. This represents a major difference with respect
to the general relativistic picture and, as will be discussed later
on, makes the parameter space for the background models effectively
unbounded.

	We next proceed to consider the eigenvalue problem and, as we did
for the general relativistic case, we neglect perturbations in the
gravitational potential (Cowling approximation) and introduce
perturbations in the velocity and pressure with a harmonic time
dependence of the type $\sim {\rm exp(-i}\sigma t)$. To simplify the
notation, hereafter we will omit the ``tilde'' on all variables, which
however should be meant to refer to normalized quantities. As a result,
the perturbed Euler and continuity equations can be written as
\begin{eqnarray}
\label{nwtpert1}
\sigma \delta {\Sigma} +
	\frac{1}{{\varpi}}
	\left({\varpi}{\Sigma}\delta
	{U}\right)' = 0\ , 
\\ \nonumber \\ 
\label{nwtpert2}
\sigma \delta {U} + 
	2{\varpi}{\Omega}\delta {W}
	-\delta {Q}' = 0\ ,
\\ \nonumber \\ 
\label{nwtpert3}
\left({\Omega}'+
	\frac{2}{{\varpi}}{\Omega}\right)
	\delta {U} + \sigma \delta {W} =0\ , 
\end{eqnarray}
where, in analogy with the corresponding relativistic quantities, $\delta
{U}$ and $\delta {W}$ are the vertically averaged radial and azimuthal
velocities, respectively ({\it cf.} eqs. \ref{vert_g1} and
\ref{var_disp}).

\begin{figure}
\centering
\includegraphics[angle=0,width=8.5cm]{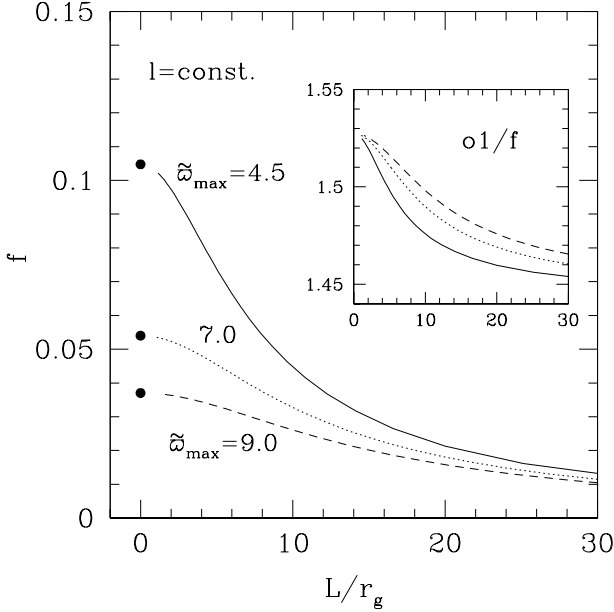}
\caption{Eigenfrequencies of the fundamental mode of oscillation for
Newtonian tori with constant angular momentum distribution orbiting a
central mass $M=10\,M_{\odot}$. We show with solid circles the values of
the Newtonian Keplerian radial epicyclic frequency at the different
locations of $\varpi_{\rm max}$, while the inset shows the ratio of the
first overtone and of fundamental eigenfrequency as a function of the
radial extents $L$.}
\label{newt} 
\end{figure}

	After some simple algebra, equations
(\ref{nwtpert1})--(\ref{nwtpert3}) are cast into a single, second-order
ordinary differential equation
\begin{equation}
\label{Q''}
\delta {Q}^{''} +
	\left[\frac{1}{{\varpi}}+
	\frac{{Q}'}{{Q}}N+
	\ln\left(\frac{\sigma}{D}\right)'\right]\delta {Q}' + 
	\frac{ND}{{Q}}\delta {Q} =
	0\ ,
\end{equation}
for the quantity $\delta {Q}$ defined as
\begin{equation}
\label{tilde_dq}
\delta {Q} \equiv \left( \frac{N}{N+1}\right) 
	\frac{\delta {P}}{{\Sigma}}\ .
\end{equation}
In equation (\ref{Q''}), the quantity
\begin{equation}
D\equiv {\sigma}^{2}- 2{\Omega} (2{\Omega} + {\varpi} {\Omega}')
	\ ,
\end{equation}
measures the deviation of the eigenfrequency $\sigma$ from the Newtonian
radial epicyclic frequency $(\kappa^2_{\rm r})_{_{\rm Newt}} \equiv
2{\Omega} (2{\Omega} + {\varpi} {\Omega}')$, which coincides with the
orbital frequency when the orbital motion is Keplerian, {\it i.e.}
$(\kappa_{\rm r})_{_{\rm Newt}} = \Omega_{_{\rm Kep}}$. Note that while
equation (\ref{Q''}) appears to be singular at the radial position where
$D=0$, it is actually regular everywhere since $\delta Q'$ and $D$ vanish
at the same position ({\it cf.} eqs. \ref{nwtpert2} and \ref{nwtpert3}).

	The eigenfrequencies and eigenfunctions of the axisymmetric $p$
modes described by equations (\ref{nwtpert1})--(\ref{nwtpert3}) have been
computed using the same numerical method discussed in the previous
Sections and some representative results are presented in
Fig.~\ref{newt}.  More specifically, we show the eigenfrequencies of the
fundamental mode of oscillation for three sequences of constant angular
momentum Newtonian tori orbiting a central source of gravitational
potential with a mass of $10 M_{\odot}$. The sequences have been built
keeping the position of the density maximum fixed ({\it i.e.}
${\widetilde \varpi}_{\rm max}=4.5$, 7.0 and 9.0, respectively) and
varying the position of the inner radius to produce tori of different
radial sizes. The inset, on the other hand, shows the ratio of the first
overtone to the fundamental eigenfrequency as a function of the 
radial size of the torus for the three different sequences. A first glance at
Fig.~\ref{newt} and a comparison with Figs.~\ref{a_fixed},
\ref{wmax_fixed} and \ref{o1_lconst}, shows that no major {\it
qualitative} differences emerge in a Newtonian framework and that all of
the most important features of $p$ modes in relativistic tori remain
unchanged also in the corresponding Newtonian models. Most notably, the
fundamental eigenfrequencies depend on the position of the mass density
and on the radial sizes of the discs, increasing as the latter
decrease. In addition, as the radial sizes tend to zero, the fundamental
eigenfrequencies tend to the values of the radial epicyclic frequencies
at the position of the mass density maxima. Finally, along all sequences,
the first and second eigenfrequencies are in a harmonic sequence 2:3 of
small integers.

	Besides this qualitative agreement, however, {\it quantitative}
differences are also present and can be easily appreciated by comparing,
for instance, the curve for $a/M=0$ in the left panel of
Fig.~\ref{wmax_fixed} with the dotted line in Fig.~\ref{newt}. Indeed,
the simultaneous presence of qualitative similarities and quantitative
differences between Newtonian and general relativistic models is not
surprising and reflects the fact that the $p$ modes are fundamental modes
of oscillation of orbiting fluid objects. As such, they depend only in
part on the gravitational field (or background spacetime) in which the
fluid is moving. Indeed, also for rotating stars, the qualitative
features of the $p$-mode oscillations are preserved when going from a
Newtonian description to a full general relativistic one, with quantitative
differences being present, however, and depending on the strength of the
spacetime curvature.

	Finally, note that because the radial epicyclic frequencies
represent the asymptotic limit of the fundamental eigenfrequencies for
vanishing torus sizes and since the Newtonian epicyclic frequency is
always {\it larger} than the corresponding general relativistic one for
all values of the black hole spin ({\it cf.} solid and dotted lines in
Fig.~\ref{epic}), all of the Newtonian frequencies along a sequence with
$\ell={\rm const.}$ will be larger than the corresponding relativistic
ones. This, together with the fact that the Newtonian epicyclic frequency
diverges for vanishingly small radii [{\it i.e.}  $(\kappa_{\rm
r})_{_{\rm Newt}} = \Omega_{_{\rm Kep}} \sim {\widetilde
\varpi}^{-3/2}$], makes it clear that for any observed QPO frequency it
will be possible to find a Newtonian torus of size $L$ that could produce
$p$ mode oscillations with the correct frequencies and in a 2:3
ratio. Stated differently, while the parameter space in the $(f,L)$ plane
for relativistic tori is bounded above by the sequence approaching the
largest epicyclic frequency admissible by a Kerr black hole with spin
$a/M=1$, this restriction does not exist for Newtonian tori and a
Newtonian physicist, who ignores the existence of marginally stable
orbits or of an event horizon, would be able to find an equilibrium
solution at every point in that plane.

	As a result, and at least in principle, Newtonian physics could
explain the HFQPOs if the only information available is the mass of the
central object $M$ and the frequencies of the modulation in the X-ray
luminosity $\sigma$. The only way to remove this ambiguity and thus
disprove the interpretation of the Newtonian physicist, would be to
make combined measurements of $M$ and $\sigma$, together with the radial
extension of the torus $L$, or of the position of the mass density
maximum $\varpi_{\rm max}$. While both these observations wwould probably
be difficult to make in practice, similar difficulties are found also in
proving the existence of stellar-mass black holes in X-ray binaries
(Abramowicz, Klu\'zniak and Lasota, 2002). This is simply the consequence
of the fact that no limit exists in Newtonian physics to the compactness
of the sources of gravitational potential.

\section{Conclusions}
\label{conclusions}

	We have presented the study of the oscillation properties of
non-selfgravitating tori orbiting around Kerr black holes. This work
extends the previous investigations of the oscillation properties of
relativistic tori in a Schwarzschild background and their importance in
explaining the HFQPOs in X-ray binaries containing a black hole candidate
(Rezzolla et al., 2003a).  Following the same approach presented in these
previous works, we have here considered the axisymmetric $p-$mode
oscillations of relativistic tori assuming a vertically integrated
description to reduce the eigenvalue problem to the solution of a system
of coupled ordinary differential equations.

	We have first considered a local analysis in a Kerr spacetime and
determined the relations between acoustic and inertial waves, showing
that both are present in the oscillations of geometrically thick discs
and that they play a different role depending on the radial size of the
disc and on the position of the rest-mass density maximum. We have then
computed the eigenfunctions and the eigenfrequencies of the axisymmetric
$p$ modes for a large variety of background models which differed either
in the spin of the black hole or in the distributions of specific angular
momentum considered. The latter, in particular, have been considered to
be constant or to increase outwards following a power-law in radius.

	On the whole, the $p$-mode oscillations of vertically integrated
tori in a Kerr spacetime possess all of the most important features
already encountered in a Schwarzschild spacetime. Firstly, the
fundamental eigenfrequencies depend on the position of the rest-mass
density maximum and on the radial size of the discs, increasing as the
latter decreases. Secondly, the fundamental eigenfrequencies tend to the
values of the radial epicyclic frequencies at the position of the mass
density maxima as the radial sizes of the tori tend to zero. Finally, for
tori constructed with constant distributions of specific angular
momentum, the first and second eigenfrequencies are in a harmonic
sequence 2:3 of small integers. This ratio is not exact but is very
accurately satisfied with deviations of $\sim 10\%$ at most. For
non-constant distributions of angular momentum, on the other hand, the
harmonic sequence is still present for sufficiently large tori, but the
deviation from a precise 2:3 ratio increases, being $\sim 20\%$ at most.

	We have also investigated the implications that $p$-mode
oscillations could have on the HFQPOs observed in X-ray binaries
containing a black hole candidate. In a model recently proposed, the
presence of a sub-Keplerian flow ({\it i.e.} a torus) near the black hole
and its oscillations are used to explain many of the features of the
HFQPOs and, in particular, the 2:3 ratio in the peaks of the power
spectral density (Rezzolla et al., 2003a). In this model, the measure of
the oscillation frequencies and the knowledge of the $p$-mode properties
can be used to measure directly the black hole properties, such as the
mass and spin, once a reasonable estimate of the torus size is made.

	In view of this, we have considered whether an equivalent
interpretation of the HFQPOs phenomenology can also be made in terms of a
purely Newtonian description of physics. To this scope, we have performed
a global analysis of vertically integrated Newtonian tori following the
same mathematical and numerical approach developed for tori in a Kerr
spacetime. Our results indicate that, at least {\it qualitatively}, the
$p$-mode oscillations of Newtonian tori preserve all of the properties
encountered in their relativistic counterparts. Furthermore, while {\it
quantitative} differences do exist between the two frameworks, it may be
difficult to distinguish between the two unless the oscillation
frequencies are measured together with the radial extension of the torus
$L$, or through the measurement of the position of the mass density
maximum $\varpi_{\rm max}$.

\section*{Acknowledgments}

	It is a pleasure to thank Marek Abramowicz, Omar Blaes, Toni
Font, and Wlodek Klu\'zniak for many useful discussions. We are also
grateful to John Miller and Tom Maccarone for carefully reading the manuscript and to
Olindo Zanotti for a close comparison of our results with his numerical
simulations. Financial support has been provided by the MIUR and INFN
(OG51). SY is supported by the NSF grant PHY 0071044 and acknowledges
support from the INFN while visiting SISSA. The computations were
performed on the Beowulf Cluster for numerical relativity {\it
``Albert100''}, at the University of Parma.

\appendix
\section[]{On the dispersion relation for tori around Kerr black holes}
\label{appendix}

	We here briefly discuss the simplifications of equations
(\ref{euler-rad})--(\ref{cont_gr}) that lead to the homogeneous linear
system with matrix form (\ref{lin_sys}). In particular, we start from the
analogy of equation (\ref{euler-rad}) with its Newtonian counterpart,
equation (\ref{nwtpert2}), whose pressure-gradient term $\delta Q'$ and
centrifugal term $2\varpi\Omega\delta W$ are of about the same order.
The reason for this is that the inertial-acoustic modes are the results
of small deviations from the equilibrium between the centrifugal force
and the pressure gradients, which produce the small perturbative radial
velocity field expressed by the first term in equation (\ref{nwtpert2}).

	We expect this to be true also in the general relativistic case
and assume, therefore, that the second and the third terms in equation
(\ref{euler-rad}) are of the same order. We also note that since
$u^t={\cal O}(1)$, then $\alpha={\cal O}(1)$ and thus the leading order
coefficient in the square bracket multiplying $\delta W$ is the first
one, {\it i.e.}  $(\Delta^{3/2}/A) (A/\varpi^2)' \Omega$.

	Because now $\Delta\sim \varpi^2$ and $A\sim\varpi^4$, it is not
difficult to show that the ratio of the third to the first term in
equation (\ref{euler-rad}) is given by
\begin{equation}
\label{A1}
	\frac{\alpha({\Delta}/{\varpi^2})\delta Q'}
	{({\Delta^{3/2}}/{A})({A}/{\varpi^2})'\Omega
	\delta W}
	\sim \frac{k}{\Omega} \frac{\delta Q}{\delta W} \ , 
\end{equation}
where we have replaced radial derivative with a factor ${\rm i}k$. Since
we assume the ratio (\ref{A1}) to be ${\cal O}(1)$, we can deduce that
\begin{equation}
\label{A2}
\frac{\delta W}{\delta Q} \sim \frac{k}{\Omega}  
	\ . 
\end{equation}

	Using now this order of magnitude estimate, we can consider the
azimuthal Euler equation (\ref{euler-phi}) and evaluate the ratio between
the first and third terms, which is given by
\begin{equation}
	\frac{({\sigma\varpi^2\sqrt{\Delta}}/{A})\delta W}
	{[{A\sigma\alpha(\omega-\Omega)}/{\Delta\varpi^2}]\delta Q}
	\sim \frac{1}{\Omega}\frac{\delta W}{\delta Q}
	\sim \frac{k}{\Omega^2} \ ,
	\end{equation}
As a result, in the limit of large wavenumbers considered here ({\it
i.e.} for $k\to\infty$), the third term in equation (\ref{euler-phi}) can
be neglected when compared with the first one.

	Finally, in the continuity equation (\ref{cont_gr}), we note that
all of the terms except the first one are multiplied by
$\widetilde{\Gamma}$ or comparable terms [{\it e.g.} $P'/(E+P)$]. The
latter is rather small and of the order $\sim (c_s/c)^2$, where $c_s$ and
$c$ are local sound speed and the speed of light, respectively. As a
result, the leading terms we retain are the first and the second one, for
which the radial derivative introduces a factor ${\rm i}k$ balancing the
smallness of $\widetilde{\Gamma}$.

\label{lastpage}  

\end{document}